\documentclass[]{aa}  

\usepackage{graphicx}
\usepackage{float}
\usepackage{txfonts}

\usepackage{placeins}
\usepackage{makecell}
\usepackage{multirow}
\usepackage[backref]{hyperref}
\usepackage{tabularx}
\usepackage{siunitx}
\usepackage{multirow, booktabs}
\usepackage{url}
\usepackage{amssymb}
\usepackage[]{units}
\usepackage[switch]{lineno}
\usepackage{changes}

\makeatletter
\renewcommand*\aa@pageof{, page \thepage{} of \pageref*{LastPage}}
\makeatother

\begin{document}
\title{The H.E.S.S. transients follow-up system}
    \author{C. Hoischen \inst{1}
    \and M. F\"u{\ss}ling \inst{2}
    \and S. Ohm \inst{2}
    \and A. Balzer \inst{3}
    \and H. Ashkar \inst{4}
    \and K. Bernl\"ohr \inst{5}
    \and P. Hofverberg \inst{5,\thanks{now at Centre Antoine Lacassagne, Nice, France}}
    \and T. L. Holch \inst{2}
    \and T. Murach \inst{2}
    \and H. Prokoph \inst{2}
    \and F. Sch\"ussler \inst{4}
    \and S. J. Zhu \inst{2}
    \and D. Berge \inst{2}
    \and K. Egberts \inst{1}
    \and C. Stegmann \inst{2,1}}
    
    \institute{Institut f\"ur Physik und Astronomie, Universit\"at Potsdam,  Karl-Liebknecht-Strasse 24/25, D 14476 Potsdam, Germany 
    \and Deutsches Elektronen-Synchrotron DESY, Platanenallee 6, 15738, Germany
    \and GRAPPA, Anton Pannekoek Institute for Astronomy, University of Amsterdam,  Science Park 904, 1098 XH Amsterdam, The Netherlands
    \and IRFU, CEA, Université Paris-Saclay, F-91191 Gif-sur-Yvette, France
    \and Max-Planck-Institut f\"ur Kernphysik, P.O. Box 103980, D 69029 Heidelberg, Germany\\
    \email{clemens.hoischen@uni-potsdam.de, matthias.fuessling@desy.de, stefan.ohm@desy.de}}
\date{Received January 7, 2022; accepted XX YY, 2022}

\abstract{Observations of astrophysical transients have brought many novel discoveries and provided new insights into physical processes at work under extreme conditions in the Universe. Multi-wavelength and multi-messenger observations of variable objects require dedicated procedures and follow-up systems capable of digesting and reacting to external alerts to execute coordinated follow-up campaigns. The main functions of such follow-up systems are the processing, filtering and ranking of the incoming alerts, the fully automated rapid execution of the observations according to an observation strategy tailored to the instrument, and real-time data analysis with feedback to the operators and other instruments. H.E.S.S. has been searching for transient phenomena since its inauguration in 2003. In this paper, we describe the transients follow-up system of H.E.S.S. which became operational in 2016. The system allows H.E.S.S. to conduct a more versatile, optimised and largely autonomous transient follow-up program, combining all major functionalities in one systematic approach. We describe the design, central functionalities and interfaces of the follow-up system in general and its three main components in detail: the Target of Opportunity (ToO) alert system, the data acquisition and central control system, and the real-time analysis. We highlight architectural decisions and features that enable fully automatic ToO follow-up and indicate key performance metrics of the sub-systems. We discuss the system's capabilities and highlight the need for a fine-tuned interplay of the different sub-systems in order to react quickly and reliably. Lessons learnt from the development, integration and operation of the follow-up system are reviewed in light of new and large science infrastructures and associated challenges in this exciting new era of inter-operable astronomy.}

\maketitle

\section{Introduction}

Dynamical astrophysical processes manifest themselves in time-variable emission of electromagnetic radiation, gravitational waves and/or particles such as neutrinos or cosmic rays. The detection and characterisation of variable or transient astrophysical sources via the different messengers hence allows us to study the underlying physics in often extreme environments. Transient objects have been studied across the electromagnetic spectrum over timescales ranging from microseconds to tens of years and comprise a multitude of different phenomena such as Fast Radio Bursts (FRBs, $10^{-6}\,s$), Gamma-Ray Bursts (GRBs, $(1 - 100)\,s$), flaring Active Galactic Nuclei (AGN, minutes -- months), or Supernovae (days -- months). Recently, the detection of high-energy astrophysical neutrinos and gravitational waves in coincidence with electromagnetic counterparts opened the window into multi-messenger time-domain astronomy.

The major challenge in time-domain astronomy lies in the unpredictable nature of variable and transient sources. They either require monitoring campaigns targeting known objects, or instruments with a large field-of-view (FoV) that are able to cover major parts of the sky to serendipitously observe e.g. explosive events. Such telescopes are operated on Earth or on satellites in space across the electromagnetic spectrum from radio, optical, to X-ray wavelengths and up to the gamma-ray regime. Also cosmic-ray, gravitational wave, and neutrino telescopes simultaneously observe major parts of the sky.

The ability to detect astrophysical transients of varying duration is directly linked to the accessible FoV of instruments, their sensitivity in the respective wavelength band, and the time to react to triggers and speed to re-point to a new sky position for pointed instruments. Depending on the science focus of instruments in the different wavelength ranges, the designs of telescopes vary. They can be broadly grouped into two categories: {\it Pointed} instruments with FoVs of at most a few tens of square degrees typically cover less than 1\% of the accessible sky. {\it Survey} instruments, on the other hand, are designed to cover major parts of the sky, typically between 10\% (e.g. {\it Swift}-BAT) to $\sim$50\% (e.g. {\it Fermi}-GBM) or truly all-sky (e.g GECAM, LVO, or IceCube). The detection of astrophysical transients with pointed instruments is either realised by serendipitous discoveries or by targeting the transients detected by survey instruments in a follow-up observation. Imaging Atmospheric Cherenkov Telescopes (IACTs) are pointed instruments with FoVs of $<$20 square degrees and are sensitive in the tens of GeV to TeV gamma-ray energy range. Their limited FoV results in an improved sensitivity compared to survey instruments --- ideal to follow-up on external triggers.

Fast transients with a duration of seconds to minutes, such as Gamma-ray Bursts (GRBs), require a swift communication of their detection and main characteristics to the astrophysical community. Survey instruments analyse the observations in real-time, detect a transient in the FoV, classify it, and send this information to relay stations across the globe from where information is distributed further via international networks. Follow-up instruments then receive and process this information and decide if they {\it can} and {\it want to} react, based on the science case, taking into account the observing conditions, and then initiate follow-up observations.

The classification and distribution of transient alerts is mostly realised by the instruments that search for them. For decades, the distribution of alerts in the community has been accomplished through human-generated Astronomer's Telegrams~\citep{rutledge1998astronomer} or via the Gamma-Ray Coordinates Network \citep[GCN;][]{barthelmy2008gcn} and streams of machine-readable notices to which observatories can subscribe.
With an increased interest in the time domain community, more and more functionality, such as advanced brokering, additional processing as well as the generation of new alerts from combinations of data streams, is sought after. A growing number of systems aim to realise these features, of which H.E.S.S. currently uses (i) AMON~\citep{AMON2013}, which combines sub-threshold data streams to generate new alerts that are sent to the GCN; (ii) AMPEL~\citep{AMPEL2019}, which provides an open source framework to combine, filter and process data across the electromagnetic-spectrum and astrophysical messengers to classify different kinds of transients from which H.E.S.S. is receiving Nova candidates; (iii) Astro-COLIBRI~\citep{astro_colibri} that allows for visual monitoring of transient alert localisations and timelines which many H.E.S.S. Target-of-Opportunity (ToO) PIs use to monitor recent alerts; or (iv) FLaapLUC~\citep{lenain2018flaapluc} which continuously runs likelihood {\it Fermi}-LAT data analyses of potentially variable sources in order to trigger follow-up observations manually.

The follow-up of transient alerts by pointed instruments requires dedicated and integrated transients follow-up systems that connect external alerts to the instrument. They are optimised for a fast and flexible handling of external transient alerts, including: alert reception and processing, execution of follow-up observations, real-time analysis (RTA) of incoming data, and communication of results and follow-up alerts to the operators, principal investigators (PIs) and the scientific community. 

In this work, we present the transients follow-up system of the H.E.S.S. experiment, an IACT system situated in Namibia, probing the very-high-energy (VHE, 30\,GeV to 100\,TeV) gamma-ray sky. The overall concept, design and main building blocks of the system will be presented in section~\ref{sec:system_properties}. The ToO alert system, the Data Acquisition and Central Control (DAQ) ToO system, and the RTA are presented in sections \ref{sec:transients_handling}, \ref{sec:daq} and \ref{sec:RTA}, respectively. The interplay of the different components and application to science cases will be discussed in section~\ref{sec:realworld}. A summary and outlook will be given in section~\ref{sec:summary}.

\section{The transients follow-up system in H.E.S.S.}
\label{sec:system_properties}

\begin{figure*}[t!]
\centering
\resizebox{0.75\textwidth}{!}{\includegraphics{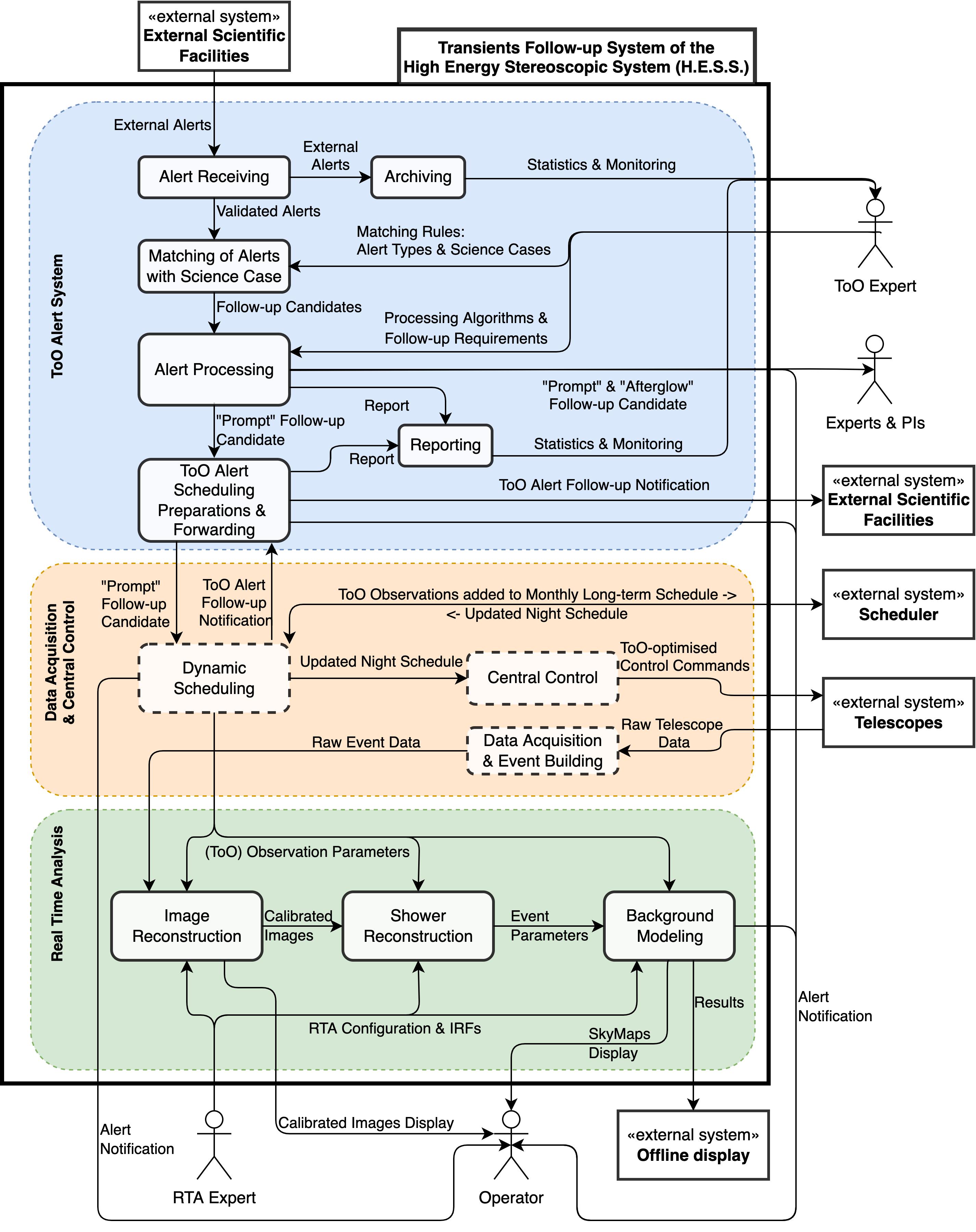}}
\caption{Functional view of the transients follow-up system of H.E.S.S. The main subsystems are the ToO alert system, the DAQ and the RTA. The three subsystems are encoded by the blue, orange and green colour areas, respectively. The main functions of each sub-system are depicted as well as the interfaces to external systems and experts and operators.}
\label{fig:too_system_overview}
\end{figure*}

\subsection{The High Energy Stereoscopic System}
H.E.S.S. is an array of five IACTs located in the Khomas Highland in Namibia and operating during clear nights with moon fraction below 40\%, resulting in a duty cycle of $(10-15)\%$. The telescopes are equipped with photomultiplier-tube-based pixelated cameras that record Cherenkov light emitted by extensive air showers, which are initiated when VHE gamma rays, electrons, and charged nuclei interact in the atmosphere. H.E.S.S. consists of two different telescope types: 4 telescopes (CT1--4) have a FoV of 5$^{\circ}$ diameter and a mirror area of 107\,m$^{2}$. CT1--4 are arranged in a square with side-lengths of 120\,m. A larger telescope, CT5, with a mirror area of 614\,m$^{2}$ and a camera FoV of $3.4^{\circ}$ diameter, is located in the centre of the array. While CT1--4 detect gamma rays with energies of $\gtrsim$100\,GeV, CT5 measures gamma rays with energies of $\gtrsim$30\,GeV. This low energy threshold is key when studying transients, in particular those that are of extragalactic origin. High-energy and VHE photons with energies beyond $\sim$100\,GeV increasingly suffer from photon-photon absorption in the extragalactic background light on their way from the source to Earth. For instance, at 100\,GeV photon energies, the gamma-ray horizon is limited to sources at redshifts of $\lesssim$1. The energy threshold of IACTs increases with e.g. increasing zenith angle of observations, limiting the reach in volume to astrophysical transients further. On the other hand, the response of the CT5 telescope and the DAQ have been optimised for a very fast re-positioning to capture short-lived transients that are expected to dim rapidly in gamma rays~\citep{Hofverberg2013,balzer2015performance}. Situated in the Southern Hemisphere, H.E.S.S. is the only instrument that can provide follow-up observations in the VHE gamma-ray band for southern-sky transients. For transients on hour timescales, H.E.S.S. has a $>$20 times better energy flux sensitivity in its core energy range above 300\,GeV than the space-based {\it Fermi}-LAT has at 1\,GeV energies\footnote{\url{https://fermi.gsfc.nasa.gov/ssc/data/analysis/documentation/Cicerone/Cicerone_LAT_IRFs/LAT_sensitivity.html}} \citep[][]{HESS_GRB2021}.

H.E.S.S. is conducting follow-up observations of transients since its inauguration in 2003 \citep[see e.g.][]{aharonian2009hess}. Dedicated working groups in the collaboration are supporting multi-wavelength efforts and decide on follow-up strategies. This work focuses on the technical developments and features of the automated transients follow-up system that executes the various programs and strategies.

\subsection{Functionalities of a transients follow-up system}
The transients follow-up system connects H.E.S.S. with ground- and space-based telescopes and transients data networks such as GCN. Based on the type and properties of newly detected astrophysical transients, the system is designed to dynamically schedule and execute automatic follow-up observations that are tailored to pre-defined science cases. As such, the system has to be able to digest diverse and limited external alert information, and to consider the visibility of objects as well as the ranking of follow-up observations against ongoing or scheduled observations. It is crucial that all parts of the system are smoothly connected, from the reception of alerts, to the reaction of the telescopes, and optimised RTA of the H.E.S.S. data on site. The key functionalities necessary for a successful alert follow-up with H.E.S.S. or other IACTs are to:

a) Receive alerts via different input channels:
Satellites and ground-based telescopes disseminate transients alert information on various timescales via different channels and with varying information content. A flexible follow-up system is therefore required that receives this information and handles the different alert types and their content --- a task best addressed by utilising international standards where possible.

b) Process and rank alerts following a matching with proposal-based science cases:
 A matching of alert types and triggers from different instruments to science cases and their sub-categories is required to rank alerts and decide if a transient should be observed. The decision is based on the alert information and must take into account the duty cycle of the ground-based instrument and the visibility of the target to propose either starting observations immediately ({\it prompt mode}) or changing the observation schedule for the following nights ({\it afterglow mode}). In addition, the decision and proposed follow-up strategy may be based on information derived from more complex processing of the original alert information, e.g. to produce a {\it tiled} pointing for the observations of transients with a localisation region larger than the H.E.S.S. FoV.

c) Execute observations according to the needs of the science case:
Transients develop on a wide range of time scales, and so does the time required to gather and disseminate reliable information. While some transients are detected and communicated to partners within seconds (e.g. GRBs), other triggers are only issued after minutes to hours due to more time-consuming analyses, human vetting, or other technical reasons. Likewise, H.E.S.S. needs to react promptly or in afterglow mode, depending on the timing of the expected gamma-ray signal of the transient and of the accompanying alert. In case of a prompt follow-up, the fully automatic transition to and execution of the new observations need to be managed.

d) Provide feedback to experts, scientific community and the system itself:
Preliminary science results from the follow-up observations are produced in real time to decide whether to extend or discontinue the observations. Timely preliminary science results of the follow-up are essential for the dissemination of potential new alerts, containing these results, to external facilities.

All these functionalities are covered in the transients follow-up system of H.E.S.S., and will be discussed in the following.

\subsection{Overview of the H.E.S.S. transients follow-up system}
\label{sec:system_overview}

The general functionalities and interfaces between the different parts of the system are summarised in Fig.~\ref{fig:too_system_overview}. The majority of tasks are executed fully automatically and handled internally. Human interaction with the system mainly happens through the configuration of sub-systems, the monitoring of the system response, generation of reports, as well as vetting of delayed observations by operators.

{\bf The ToO alert system} acts as a gateway and receives external transient alerts. Upon alert reception, it evaluates the alert properties and matches them with science cases that are part of the H.E.S.S. transients science program with a number of pre-approved triggers. This task can include more complex calculations and/or the construction of an observation strategy for the follow-up of the transient. The configuration and monitoring of the ToO alert system is handled by the ToO alert system expert. Information about transient alerts that match science cases are provided to respective PIs. If a transients candidate passes all requirements for follow-up observations, it is forwarded to the DAQ for scheduling.

{\bf The DAQ ToO system} handles the scheduling of accepted follow-up candidates with H.E.S.S. and interacts with the monthly long-term schedule and the observation program planned for the current night. It receives the details of filtered and ranked ToO candidates from the ToO alert system, stops ongoing observations, initiates the (rapid) slewing of telescopes to the new position and starts the data acquisition in the region of interest. Data is recorded via the DAQ and processed in the RTA.

{\bf The RTA} performs an online analysis of the live data stream that is received from the cameras, applies online calibration methods, reconstructs the particle shower properties, and selects and displays gamma-ray-like candidate events. The results, including live sky images, are displayed to the operators and stored on a web-page and in an archive for further investigations.

The design, implementation and functionality provided by each subsystem will be introduced in the following sections.

\section{The ToO alert system}
\label{sec:transients_handling}

The ToO alert system (blue section in Fig.~\ref{fig:too_system_overview}) listens for new alerts on various data streams from external networks simultaneously and matches them with pre-defined H.E.S.S. science cases. The configurations for each science case define the processing steps and trigger criteria to filter follow-up candidates and to derive the observation strategy and alert prioritisation. More complex alert processing steps are realised in a dedicated pipeline built from modular scripts. The scheduling parameters derived during the processing are then reported to science and operation experts and, in case of automatic follow-up, passed on to the DAQ for immediate follow-up observations. A detailed view of the processing logic that is executed whenever an alert is received is given in Fig.~\ref{fig:vo_react_to_alert}. The implementation of the ToO alert system is explained in more detail in Appendix \ref{app:ToO}. 

\begin{figure*}[t!]
\begin{center}
    \resizebox{0.75\hsize}{!}{\includegraphics{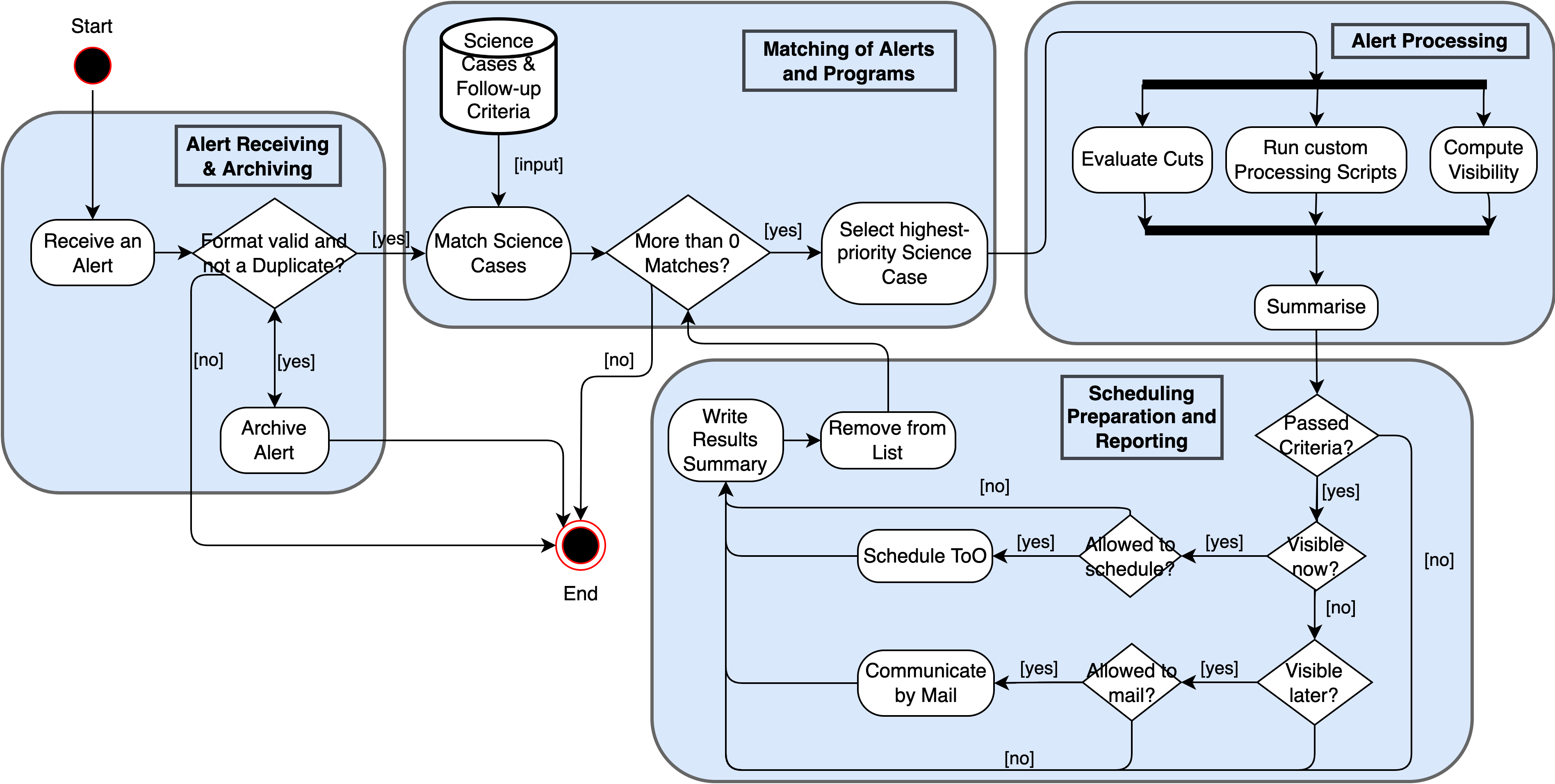}}
    \caption{Functional view of the ToO alert system as it reacts to an incoming alert. The main function groups, {\it Alert Receiving and Archiving}, {\it Matching of Alerts and Science Programs}, {\it Alert Processing} and {\it Scheduling Preparations and Reporting} are indicated by the four large boxes. The detailed functionality is described in Sec.~\ref{sec:transients_handling}. The mapping of these functions to the implementation is shown in Fig.~\ref{fig:vo_implementation_view}}
    \label{fig:vo_react_to_alert}
\end{center}
\end{figure*}

\subsection{Receiving and processing alerts}
Each transients science program is realised through a set of configurations. Each configuration applies to a single type of alert and addresses a specific science case (e.g. prompt GRB follow-up). These science case configurations define detailed aspects for the follow-up such as whether a fully automated reaction is allowed, and a time range with respect to the transient event during which this configuration should be used. Moreover, they include scheduling aspects like the maximum time to observe the target, detailed trigger criteria, or a relative ranking of each science configuration. Visibility windows are calculated for all science cases taking into account the zenith angle, sky-darkness constraints, or source properties such as reported fluxes, event counts and localisation uncertainties. This allows the system to be flexibly configured for each science case, but may require multiple updates of the schedule if updated information of a single alert is received. Furthermore, dedicated processing algorithms that e.g. match the transient's location with catalogues or perform more complex calculations are referenced by name with corresponding parameters in the configuration file. A particularly complex case is the algorithm for creating an optimised pointing pattern for the follow-up of Gravitational Wave (GW) events using the correlation of the galaxy distribution in the local Universe with the GW event localisation information, as described in~\citep{Ashkar_2021} and outlined in Sec.~\ref{sec:realworld}. This approach is also used to generate an optimised pointing pattern for large localisation uncertainties from {\it Fermi}-GBM alerts. The configuration files specify which processing tasks are executed by the pipeline. This allows one to introduce new processing functions rapidly, without changes to the overall pipeline code and behaviour. Multiple science case configurations may apply to a single alert type.

Configurations are grouped into categories with a common science goal. For instance, the category {\it GRB prompt} includes configurations for {\it Swift}-BAT, {\it Fermi}-GBM, as well as {\it Fermi}-LAT alerts. This grouping allows one to match the recipients of scheduling information to the different experts, PIs and/or entire mailing lists. The grouping mechanism is also useful to define test configurations in order to test new processing functions or monitor the system without actively triggering observations. The human-readable scheduling information is appended to a report with dedicated instructions and information based on the assigned group that is sent to the PIs by E-mail.

If all processing steps and trigger conditions are fulfilled and the observation strategy has been defined, the follow-up candidate together with the scheduling information is forwarded.

\subsection{Forwarding follow-up candidates}
The forwarding of follow-up candidates is split into two modes: the prompt-mode forwarding that initiates a fully automatic start of the observations, and the afterglow mode where experts and operators are informed about the upcoming observation opportunity. Usage of the prompt mode requires that a flag is set in the science case configuration.

The prompt mode requires that the observation window starts within a few minutes and continues for at least 5 minutes. A fast evaluation of the short-term visibility is carried out to minimise the time to initiate a follow-up. The entire observation window is evaluated only if the visibility criteria are met. In prompt mode, the essential scheduling information is collected and sent to the DAQ. Once the DAQ has received the information, the operators are notified via a pop-up window and an E-mail is sent to a predefined list of recipients. 

The afterglow mode alerts the operators with a pop-up window, but no further information is transmitted to the DAQ. The operators, experts and other recipients receive an E-mail with all information about the alert and its scheduling constraints and instructions. Since the ToO Alert System has determined that all criteria for a follow-up are fulfilled, the operators are expected to schedule the observations with the provided information. The E-mail contains a terminal command that can be used by the operators to insert the target into the observation schedule prior to the beginning of the night. In parallel, they can liaise with experts and PIs to discuss whether observations should indeed be pursued, and if so, fine-tune the timing of the ToO observations. The afterglow mode is used for alerts that can be followed up with delays of up to two days.

\subsection{Performance}
The most important performance metric for the ToO alert system is the time required to fully process an alert. This strongly depends on the complexity of processing functions and algorithms that are being executed. Therefore, we consider two cases:

The follow-up of GRB alerts is among the high-priority use cases of the system, where the visibility window is calculated and a few alert parameters are being used to evaluate, if a follow-up should occur. In such a case, the processing takes around 1--2 seconds up to the handover to the DAQ. A visualisation of the visibility window is typically created within 4 seconds, and sent together with the E-mail. In general, the experts and operators receive the E-mail with all the material in less than a minute in such cases.

The most complex of the implemented cases is the gravitational wave alert, which uses a galaxy targeting algorithm and takes significantly longer. Deriving the best first position to start observations usually takes around 30 seconds, including reading the galaxy catalogue and downloading and analysing the localisation map. Deriving a full tiling pattern can take up to 60 seconds, depending on how long and how much of the uncertainty region can be observed.

However, these processing times are not the only contribution to the overall delay of the response. There are important, non-negligible delays that are beyond the control of H.E.S.S., such as the delay between the astrophysical event and the reception of the alert. For GRBs, this delay is typically in the range of 30 seconds and is determined by the on-board detection and download link of the data \citep{Hoischen2018}. In the LIGO/Virgo observation runs O1$-$O3, the earliest gravitational wave alerts were sent with delays of the order of a few minutes after detection, with potentially larger delays depending on integration time or dependencies on other higher-order analysis results \citep[see e.g.][and references therein]{Ashkar_2021}.

Another key metric is the filtering performance of the system. The ToO alert system receives about 50000 alerts per month\footnote{The rate of alerts received is very non-uniform and strongly depends on the alert strategy of the brokers that the H.E.S.S. System is subscribed to.}. Only about 50 alerts are matched with science case configurations, which is a reduction by a factor of 1000.

The availability of the ToO alert system is another critical factor, as no alerts can be received during downtime. A watchdog for the \textit{Alert Receiver} ensures a maximum up-time. Downtime of the order of 20 seconds occurs every two months when the system is maintained and updated. Rare problems with the internet connection to the H.E.S.S site on the Khomas Highland are another source for downtime.

\subsection{Toolkit for validation, development and alerting}

Beyond the core functionality of processing received alerts, a number of useful tasks can be performed with the ToO alert system. Tools are available to generate test alerts that are sent to the \textit{Alert Receiver} to validate the entire chain of processing alerts up to the start of observations in so-called fire drills (see Sec.~\ref{sec:realworld}). Independent of other parts of the H.E.S.S. transients follow-up system, the ToO alert system software can be used in a standalone mode. Large numbers of alerts\footnote{Public databases of VOEvents, such as voeventdb.remote~\citep{staley20164}, are extremely helpful in such tasks.} can be processed offline in order to fine-tune and test new science programs, processing functions and algorithms, to simulate the system response and to support rapid development of new features.

In addition, the ToO alert system supports a fast feedback loop on whether alerts are followed by H.E.S.S. to increase multi-wavelength coverage. It generates new alerts when a ToO follow-up is inserted into the observation schedule (see also next section), comprising information about the underlying external alert and the H.E.S.S. follow-up, such as the applied science case and priority. These alerts are then sent\footnote{{\it Comet} provides such a functionality implementing all VOEvent standards.} to ATOM~\citep{hauser2004atom}, the robotic optical telescope on the H.E.S.S. site, to trigger simultaneous optical ToO observations (see Fig~\ref{fig:too_system_overview}).

\section{The DAQ ToO system}
\label{sec:daq}

\begin{figure*}[t!]
\begin{center}
    \resizebox{0.75\hsize}{!}{\includegraphics{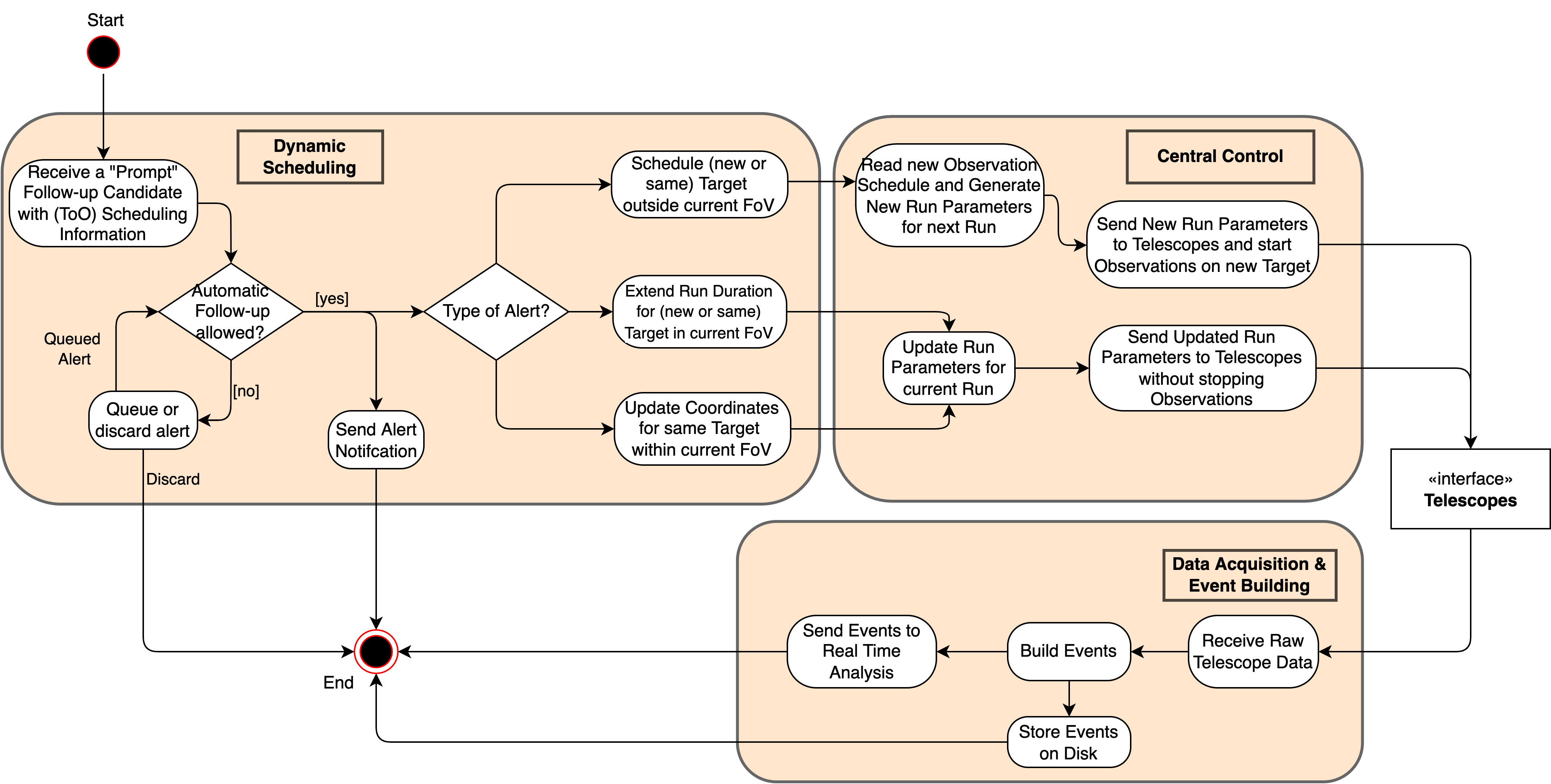}}
    \caption{Functional view of the DAQ system as it reacts to a prompt follow-up candidate. The central functions are {\it Dynamic Scheduling}, {\it Central Control} and {\it Data Acquisition and Event Building}. Further details can be found in Sec.~\ref{sec:daq}. The implementation of these functions as components is described in Appendix~\ref{app:DAQ} and shown in Fig.~\ref{fig:daq_implementation_view}.}
    \label{fig:daq_functional_view}
\end{center}
\end{figure*}

The DAQ system \citep{Balzer2014} steers all hardware and software components taking part in the observations and provides a framework for their monitoring and control, configuration, logging and error handling. The DAQ system is distributed over about 200 processes with dedicated software representations (with \textit{Managers}, \textit{Controllers}, and \textit{Readers}) of the controlled hardware and software components, a central control component organising the execution of the scheduled observation runs, and a collection of databases for the configuration of the DAQ system and the H.E.S.S. array elements. It also provides the central user interfaces in the control room through which operators interact with the H.E.S.S. array. The DAQ system runs on a computing cluster with off-the-shelf components~\citep{Zhu2021_DAQCluster} that provide the necessary computing and storage resources. In this section, we will give an overview on those components and functions of the DAQ system that have been enhanced with new features for the ToO follow-up.

An overview of the main functions and components of the DAQ system relevant for the ToO follow-up are given in Fig.~\ref{fig:daq_functional_view} and in Fig.~\ref{fig:daq_implementation_view}, respectively. As the DAQ system is responsible for the operation and integration of all other H.E.S.S. systems, it interfaces with the ToO alert system and the RTA, as well the nightly scheduler and the telescopes, as part of the transients follow-up. The three main functions related to the ToO follow-up are discussed below in relation to their ToO capabilities for prompt alerts, while the implementation of the DAQ ToO system is described in more detail in Appendix \ref{app:DAQ}.

\subsection{Dynamic Scheduling}
The DAQ system is responsible for the dynamic scheduling (see Fig.~\ref{fig:daq_functional_view}) of ToO observations during the night, in response to a follow-up candidate alert received from the ToO alert system. When the automated reaction of the system is active, different reaction schemes are invoked depending on the type of the incoming follow-up candidate and the status of the ongoing observations. If the incoming follow-up candidate can be observed immediately (\textit{prompt} mode) and has a higher priority than the ongoing observations, a fully automatic execution of the ToO observations is initiated. The type of the prompt alert leads either to the insertion of observations for the new ToO target into the schedule for the night, or to an update of the target position or duration of already ongoing observations without interrupting their execution:
\begin{itemize}
    \item if the new ToO target is located outside a 1.5$^{\circ}$ radius around the current pointing direction, i.e. close to the edge or outside the field of view of the currently ongoing observations, a new schedule including the ToO observations is prepared, the ongoing observations are stopped, and the ToO observations are started,
    \item if the new ToO target is located within the 1.5$^{\circ}$ radius, thus well within the current field of view, the duration of the ongoing observations is prolonged without restarting the observations,
    \item if the incoming alert is a position update of a previously received alert that is currently being observed, the telescopes are repointed to the new ToO target position without restarting the observations, only the data acquisition is paused for the duration of the repointing.
\end{itemize} 
In all other cases, the follow-up to the alert is delayed and involves the operators to schedule the observations (\textit{afterglow} mode). Once the prompt follow-up is initiated, the DAQ system informs all relevant stakeholders, including the operators via pop-ups, the PIs and experts via E-mail, and the ToO alert system for fast feedback to external facilities.

\subsection{Central Control}
The DAQ system's central control organises the execution of the observations and steers the telescopes and participating array components through the observation life cycle. This life cycle typically consists of the stop of the ongoing observations, the distribution of the (new or updated) observation parameters to the available telescopes, the configuration and start of the new observations, and the data taking for its duration (see Fig.~\ref{fig:daq_functional_view}). For prompt ToO observations, a special ToO observation mode is used that has been optimised for a faster stop-start cycle of the observations, thereby increasing the robustness to potential failures of individual hardware and/or software components during the cycle. In addition, the typical life cycle may omit the restarting of the observations altogether depending on the type of alert (see also previous section).

CT5 is not only the H.E.S.S. telescope with the fastest slewing speed, but can also use reverse pointing for ToO observations, so that it can reach almost any position in the sky in less than one minute \citep{Hofverberg2013}. Furthermore, some aspects of the stop and start procedures of the telescope have been parallelized to speed up the stop-start cycle. The Cherenkov camera, which during non-ToO observations waits until the telescope has reached stable tracking of the target, already starts taking data during ToO observations as soon as the target position enters the field of view of the camera. As a result, CT5 reaches the target and the start of observations significantly faster than the slower CT1--4 telescopes. The DAQ system therefore already starts collecting data for ToO observations when CT5 sends its first data and allows the CT1--4 telescopes to join later during the ongoing observations. The ToO observation mode is also more robust against possible failures of telescopes or other components. In normal observations, data acquisition is interrupted if one of the telescopes fails, allowing the operators to fix the problem and continue with the regular observation schedule. However, for time-sensitive ToO observations, all non-critical components, including the CT1--4 telescopes, are marked as optional, and failures do not abort observations. Once the underlying issues are resolved, any optional component can be added to the observations later.

\subsection{Data Acquisition and Event Building}

Once the observation starts, the DAQ system continuously processes the telescope-wise event raw data sent from the Cherenkov cameras, performing basic data integrity and quality checks, and builds them into sub-array events in a standard H.E.S.S. data format. The raw telescope data is received and processed by \textit{Event Builder} processes. Each \textit{Event Builder} process receives a chunk of data from all cameras participating in the sub-array (e.g. CT5-mono, CT1-4 stereo, CT1-5 hybrid), where it is buffered and processed. The receiving \textit{Event Builder} process is switched every few seconds to distribute the event processing load across multiple Event Builders in the computing cluster, and to support incoming data rates greater than the event processing rate of a single \textit{Event Builder}. This standard data acquisition scheme did not need to be tailored to the ToO follow-up observations and is capable of handling large amounts of data expected from even the brightest bursts observed in ToO mode (up to $\sim$kHz gamma-ray rates). This data is stored on disk of the on-site data storage for later offline processing. In addition, the events are forwarded in-memory to the RTA where the data is further calibrated and analysed (see Sec.~\ref{sec:RTA}).

\subsection{DAQ Performance}

During the nearly two decades of operation of the H.E.S.S. experiment, the DAQ system underwent several major upgrades, both to add new features and to improve its performance and stability. The most recent feature upgrade was the addition of the ToO capabilities described in this work, coupled with an improvement in the response time to alerts, one of the key factors in the follow-up of prompt alerts. As a distributed control system, the DAQ introduces some overhead for inter-process communication and exchange of status information during the steering of the processes participating in the observations. This software overhead now averages to less than 2$s$~\citep{balzer2015performance} which is insignificant compared to the total response time to a ToO alert. The overall response time during the observation execution is dominated by the stop-start-cycle of the observations and especially by the slewing of the telescopes to the ToO target position. With the fast slewing speed of CT5 and its capability to perform ToO observations with reverse pointing \citep{Hofverberg2013}, the average total response time of the H.E.S.S. array in a fully automatic reaction to prompt ToO alerts is on average less than one minute for almost any position in the sky.

Another important performance indicator is the failure rate of the prompt follow-up system caused by technical problems of the DAQ system itself, or during the interaction of the DAQ system with other components during the dynamic scheduling and subsequent stop-start life cycle of the observations. While the total loss of observation time attributed to problems with the DAQ system (e.g. due to IACT hardware problems or software errors) is less than 1\%~\citep{Balzer2014}, the unique nature of transient phenomena makes availability particularly crucial. Most failures in responding to prompt ToO follow-ups have been caused by feature upgrades of the ToO system and resulting changes in the software interface between, and/or behaviour of, the components involved. Rigorous testing of all components and their interaction during fire drills, both after upgrades and at regular intervals, has proven to be very effective at minimising these failures and detecting potential problems at an early stage (see also Sec.~\ref{sec:realworld}).

The load on the computing cluster during data acquisition is governed by the processing of the telescope raw data, the subsequent event building and the real-time analysis (see also Sec.~\ref{sec:RTA}). In the current setup, the data acquisition is distributed in 25 processes over 5 computing nodes (with an Intel Xeon Silver 4114, 2 x 10 cores with 2.2 GHz, and 96 GB RAM architecture)~\citep{Zhu2021_DAQCluster} and supports a total event data rate of $2.5 - 3.0$\,kHz on average in normal data taking with 5 telescopes with less than 10\% and 35\% utilisation of the available CPU and memory, respectively. Processing sudden bursts of data from e.g. extremely bright transient events would therefore be possible with the current DAQ system.

\section{Real-time Analysis}
\label{sec:RTA}

\begin{figure*}[t!]
\begin{center}
\resizebox{0.75\hsize}{!}{\includegraphics{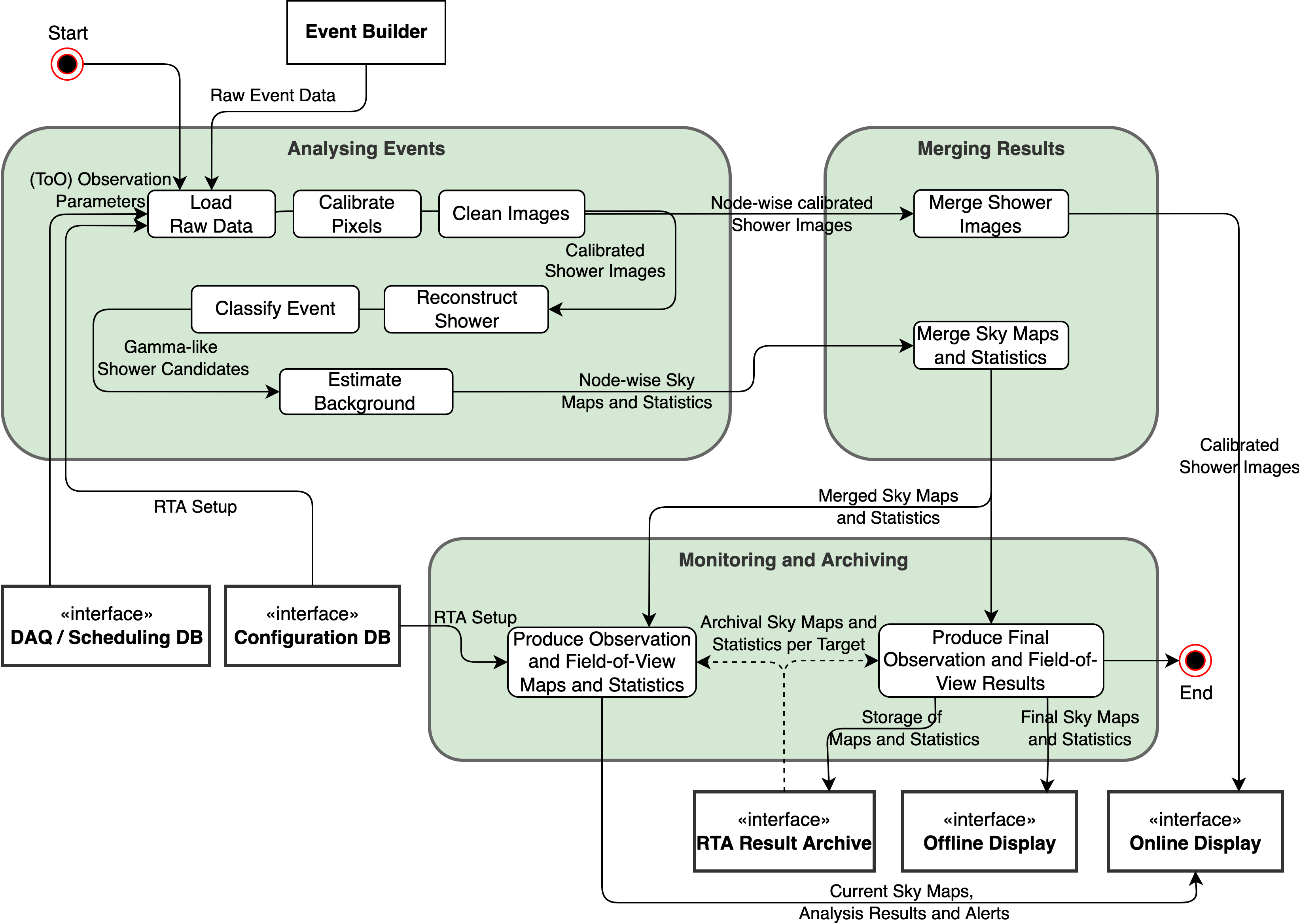}}
\caption{Functional view of the RTA system. The central functionality is described by the boxes: {\it Analysing Events}, {\it Merging Results} and {\it Monitoring and Archiving}. The details are described in Sec.~\ref{sec:RTA} and Appendix~\ref{app:RTA}.}
\label{fig:rta_func}
\end{center}
\end{figure*}

Any transients follow-up system that aims at providing feedback to array operators, other facilities, or the scientific community in general requires real-time analysis of the data being taken. The RTA in H.E.S.S. was developed for Phase I of H.E.S.S. operations, and is described in more detail in \cite{Funk2005} and \cite{Balzer2014}. In this section, we will focus on the RTA as developed for H.E.S.S. as a hybrid system of telescopes and present key performance parameters. The RTA implementation is described in more detail in Appendix~\ref{app:RTA}

Figure~\ref{fig:rta_func} shows the functional view of the RTA in H.E.S.S. It provides the following functionality:

Data access and RTA configuration: The raw events and telescope camera images are built by the \textit{Event Builder} and fed into the RTA framework. The RTA is configurable, which allows for a dynamic adjustment of RTA properties such as the \textit{thinning} of events to match computing resources. Calibration and analysis settings that set up the RTA to process incoming data with specific settings to e.g. match different sky-brightness conditions like moonless or moderate moonlight nights can also be adjusted.

Data calibration and data quality: The raw camera data is calibrated in real time, including basic identification of camera hardware problems on the pixel level, data quality monitoring and basic error handling. Cherenkov camera shower images are cleaned of night-sky-background noise for further image analysis.

Event reconstruction, classification and background modelling: Event properties, such as direction and energy of primary particles are reconstructed in real time based on the RTA-calibrated camera shower images. Fast and powerful event classification and selection of gamma-ray like events is performed based on boosted-decision trees \citep[BDTs][]{Ohm2009} and neural networks \citep{Murach2015}. Training of the classifiers is performed offline based on Monte-Carlo gamma rays and real background events, and covers the full phase space of observations (sub-arrays, zenith and offset angles, optical efficiencies).

Merging of results and background modelling: The background estimation and signal extraction in the field-of-view and the region of interest is performed per computing node, and for configurable background estimation methods. The current implementation uses the ring background technique~\citep{berge2007background}, which produces sky maps such as gamma-ray excess and significance maps. Node-wise results are then sent to a merging process, which accumulates maps and statistics.

Feedback and notification: Low-level data analysis results, such as intensity displays of telescope shower images or high-level sky maps, are shown on \textit{Online Displays} in the control room on site. During the ongoing observations, the operators are alerted via sounds and pop-up windows in case the significance at the target position exceeds configurable thresholds that are stored in the \textit{DAQ / Scheduling} database. Archival RTA results are also accumulated to the current observation providing the operators with alerts for longer observations of the same region of the sky (e.g. intra-night, over multiple nights).

Monitoring and archiving: RTA results for the individual observation run and the accumulated archival data set are stored in the \textit{RTA Result Archive} within a few minutes after an observation has concluded in form of statistics and maps in a database and on disk, respectively. They are used through a web-interface in the \textit{Offline Display} for provision to the operators, PIs of ToO programs and collaboration members. The \textit{Offline Display} provides the RTA statistics like gamma-ray excess and significance per observed target on intra-night, nightly and monthly basis. Sky maps are stored for offline handling and search for emission in the field-of-view (e.g. following updates of MWL coordinates for transient events) with more advanced methods.

\subsection{RTA Performance}
The RTA implementation described here has been running on the H.E.S.S. site in Namibia since mid-2016. During this time, the H.E.S.S. array underwent two major upgrades of the Cherenkov cameras. The original HESS-I cameras were replaced with NECTar-based readout chip design cameras in 2015 and 2016 \citep{HESS-IU}. In September 2019, the camera on the CT5 telescope was upgraded to the digital CTA-prototype FlashCam camera \citep{FC:SPIE}. In particular for the latter camera replacement, the RTA was prepared in advance of the upgrade, allowing it to detect the Crab Nebula in the first night of full operation \citep{FC:HESSNews}.

\begin{figure}[t!]
\begin{center}
\includegraphics[trim=35 0 40 0,clip,width=0.495\textwidth]{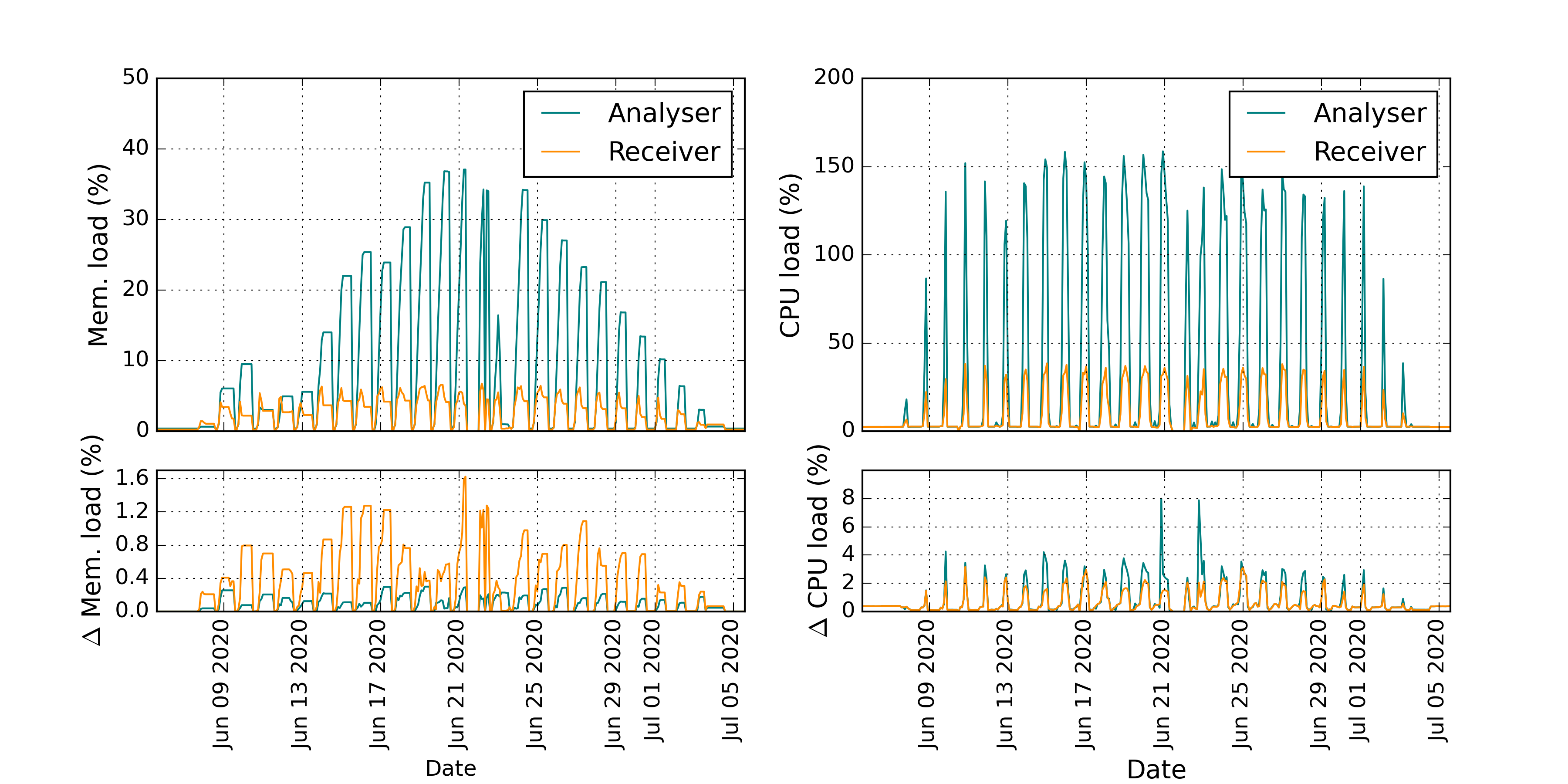}
\caption{The top panel shows the average memory (left) and CPU load (right) per night of the physical computing nodes. The bottom panel shows the standard deviation of the  memory (left) and CPU (right) load of the computing nodes. Note that 100\% CPU load corresponds to one of the 20 available cores per CPU. The {\it Analyser} process does the bulk data processing, while the {\it Receiver} process is responsible for data reception in the {\it Event Builder}.}
\label{fig:clusterperf}
\end{center}
\end{figure}

Figure~\ref{fig:clusterperf} shows the memory and CPU usage of the \textit{Analyser} and \textit{Event Builder} processes, averaged over all computing nodes, during one H.E.S.S. observing shift. The bottom panels show the standard deviation of the load over the computing nodes, and demonstrate a stable and smooth operation across cluster machines with event rates reaching 3\,kHz. Note that this metric already covers an observing month during which the CTA prototype FlashCam camera was operational in the array \citep{FC:SPIE}.

A benchmark of the \textit{Analyser} process results in the following metrics in terms of CPU needs: The data loading ($\sim$15\%), pixel calibration ($\sim$40\%) and shower image cleaning ($\sim$15\%) require the largest resources with a total of $\sim$65\%. The shower reconstruction ($\sim$5\%) and background suppression ($\sim$10\%), on the other hand, only require $\sim$15\% of the total CPU needs. Also the background modelling contributes at a lower level of $\sim$20\%.

\begin{figure}[t!]
\begin{center}
\includegraphics[trim=0 5 0 5,clip,width=0.495\textwidth]{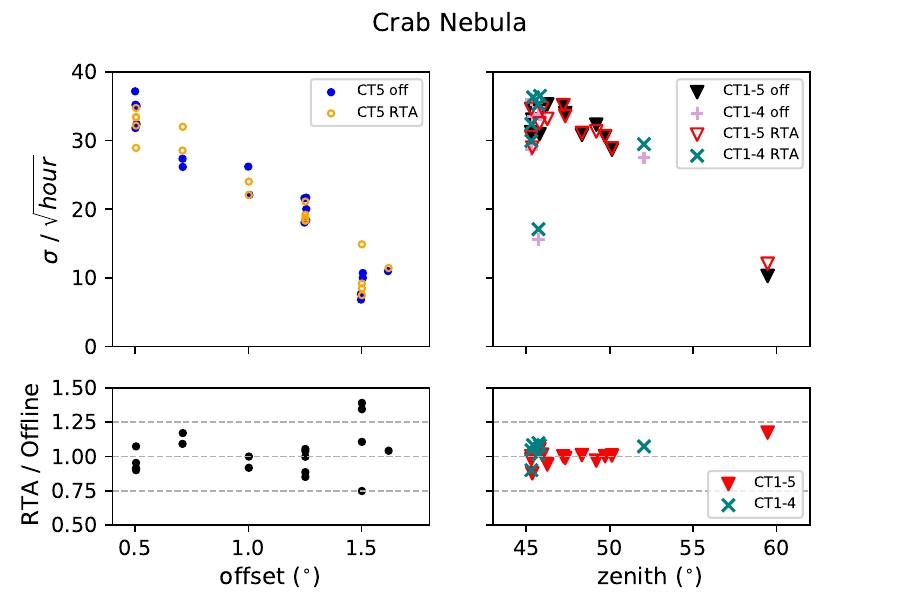}
\end{center}
\caption{Sensitivity expressed in terms of signal recorded per $\sqrt{\mathrm{time}}$ of the RTA compared to the off-site analysis using the same signal/background separation technique for Crab Nebula observations conducted under various offset (left) and zenith angles (right). The bottom panel shows the ratio between performance of the RTA and the off-site analysis. Note that bad-weather data is included, for which lower significances are reached.}
\label{fig:crab_perf}
\end{figure}

The sensitivity of the RTA has been studied by comparing the achieved RTA performance with the full off-site analysis performance, using Crab Nebula observations conducted under different observing conditions, and with different sub-arrays with data acquired between 2017 and 2019 using the previously installed CT5 camera \citep{Bolmont2014}. Fig.~\ref{fig:crab_perf} shows the significance per square root of observation time for Crab observations at various zenith and offset angles. The RTA can detect a Crab-Nebula-like gamma-ray source in less than a minute of observation time. Within a typical 30-minute observation run, sources with a strength of $\sim$5\% of the Crab Nebula can be detected. As the energy threshold increases with increasing offset to the centre of the camera, the total significance per square root hour decreases by a factor of $\sim$4 from on-axis observations to the edge of the FoV of the CT5 camera at $\sim$1.5 degrees in a CT5 mono analysis. The sensitivity of the RTA is in general comparable to the off-site analysis within 25\%. The sensitivity achieved as a function of zenith angle confirms this behaviour, and demonstrates that the RTA response is stable using default calibration coefficients and the running-pedestal estimation \citep[cf.][]{Funk2005}. A real-time evaluation of calibration coefficients was not deemed necessary, given the overall excellent RTA sensitivity achieved. The agreement between RTA and off-site analysis also suggests that both analyses achieve comparable energy thresholds. In the core energy range, between 80\,GeV and 10\,TeV, we do not expect a significant difference with the results that will be obtained using the newly installed FlashCam camera. First studies conducted with limited data sets confirm this assumption.

\begin{figure*}[t!]
\begin{center}
\resizebox{0.3425\textwidth}{!}{\includegraphics{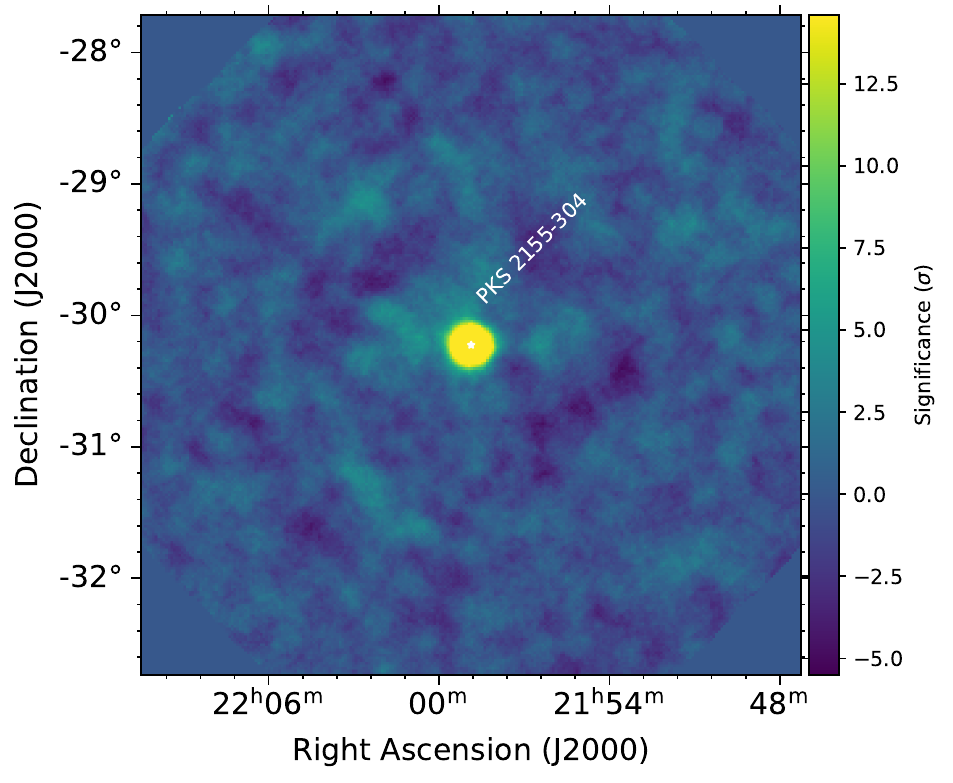}}
\resizebox{0.3335\textwidth}{!}{\includegraphics{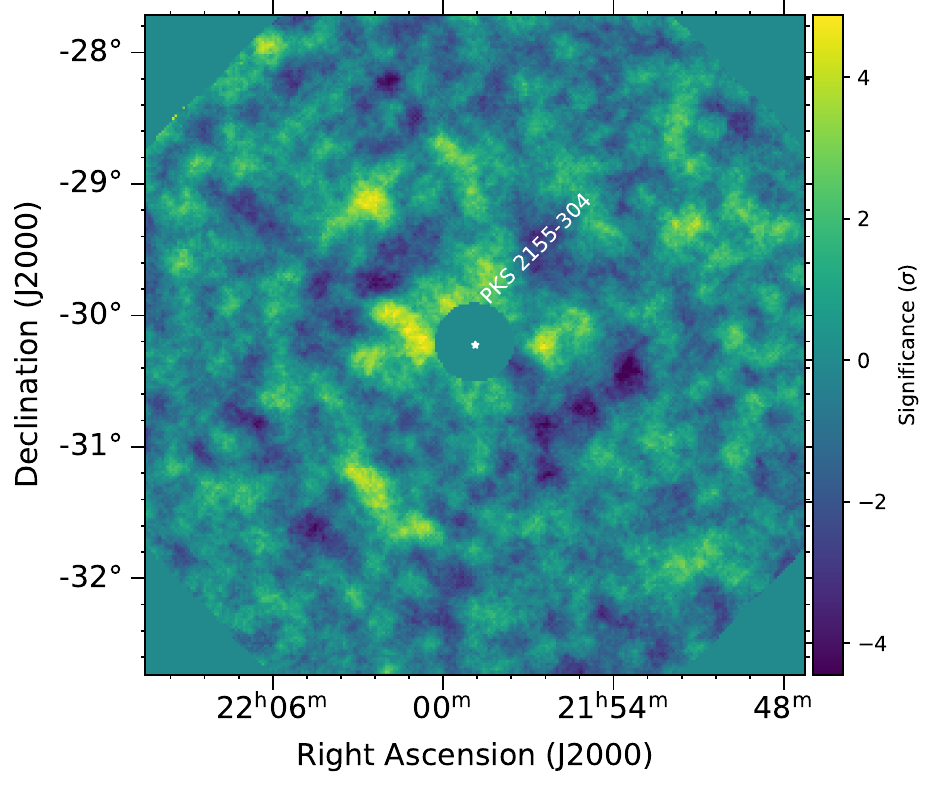}}
\resizebox{0.314\textwidth}{!}{\includegraphics{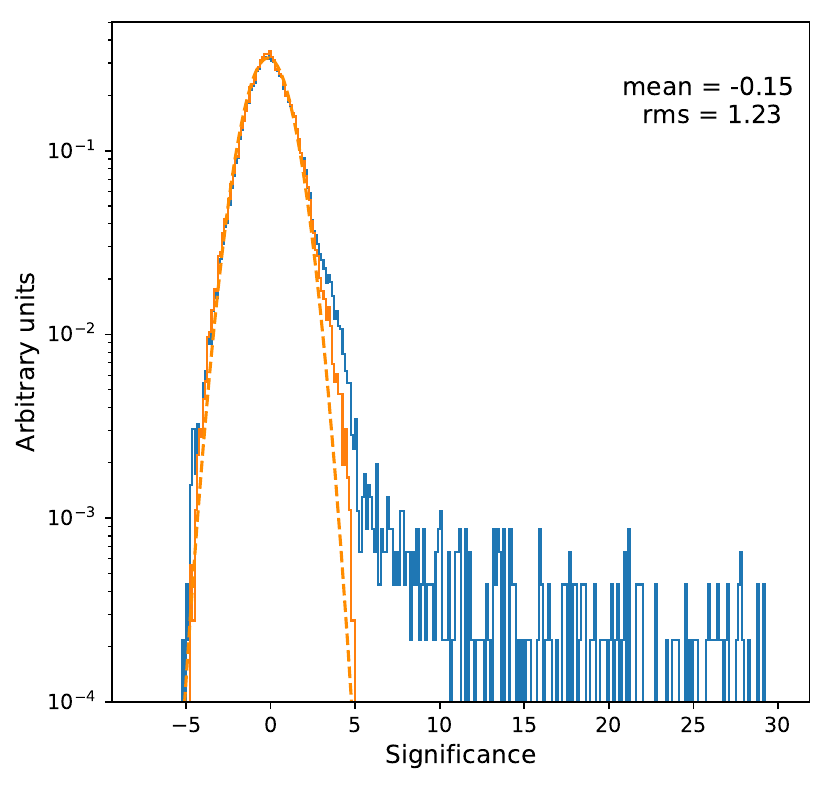}}
\end{center}
\caption{Significance map, including the source region (left), and excluding the source region (middle) for  50 hours of PKS~2155$-$304 observations conducted with the full CT1--5 array (pre-FlashCam upgrade). The right plot shows the 1D distribution of the significance (blue) and excluded significance (orange) maps. The mean and r.m.s. of a fit of a one-dimensional Gaussian function to the excluded significance distribution are also shown.}
\label{fig:rta_syst}
\end{figure*}

We studied the RTA background systematics in deep observations at the sensitivity limit of H.E.S.S. Fig.~\ref{fig:rta_syst} shows significance maps, including and excluding the source regions, as well as their corresponding 1D distributions across the FoV for a 50-hour data set taken on the blazar PKS~2155$-$304 with the full CT1--5 array. While some structures at significance levels of $\sim$4$\sigma$ are visible, the overall background is under control and no region in the sky shows significant emission beyond 5$\sigma$. Note that the RTA has to rely on observations of empty FoVs to construct the background model whereas in the off-site analysis the background model is constructed from a full observation run. This explains the somewhat increased systematics in the background estimation as apparent in the significance distributions.

\subsection{Future RTA improvements}

Further improvements to the RTA can be envisaged in several domains. The installation of more computing resources could allow us to improve the overall sensitivity of the RTA by exploiting state-of-the-art pixel-wise likelihood analysis techniques. However, the expected sensitivity improvement of $\lesssim$25\% is rather moderate, requiring a careful consideration whether the additional investment in on-site computing hardware resources is justified for the science cases at hand. The background systematics for deep observations could be further reduced by using more realistic calibration coefficients and/or by matching the background models to the calibration procedure on site (at the moment, the off-site calibration and classification is assumed for the on-site background model). Furthermore, more advanced data quality checks and even corrections could be implemented to further improve the detection and treatment of hardware defects in the RTA. Employing pixel-wise likelihood techniques or pixel-based deep-learning methods \citep{Shilon2019, Steppa2019, Parsons2020} in the RTA will require a careful assessment of e.g. the impact of turned-off or broken pixels, as well as differences between calibration coefficients used in the RTA and those assumed in the deep-learning method.

Currently, sky maps of significance and gamma-ray excess, as well as distributions of excess events as a function of squared distance to the source of interest, are displayed to the on-site observers. It would also be straightforward to implement preliminary gamma-ray source flux estimates and implement features to derive longer-term gamma-ray light curves that can be correlated with long-term light curves of sources of interest in other wavelength bands. Furthermore, at the moment, sky maps shown to the observers are updated every 30 seconds to 1 minute, while RTA results stored in the archive only capture the integrated maps over the duration of an observing run. Storing additionally the gamma-ray candidate event list would allow one to re-process the RTA results offline and to implement additional features such as the search for shorter-term transients with tools such as the ones used in the H.E.S.S. Extragalactic Survey \citep[HEGS][]{Bonnefoy:HEGS, brun2020analysis}. Another future extension could be envisaged adopting a FoV-wide search for serendipitously detected transients or a combination of RTA results across larger sky areas to search for gamma-ray emission from transients with poor localisation as discussed in the next section.

Based on pre-defined trigger criteria and for certain transients science cases, the H.E.S.S. on-site operators are instructed to monitor the RTA output while data is being taken and to manually prolong observations. A link of the RTA to the ToO alert system is envisaged and currently explored in H.E.S.S. An automatic prolongation of ongoing observations could be realised via an interface between the ToO alert system and the RTA. For instance, the RTA could send observation results shortly before the end of an observation run to the ToO alert system in the form of an (VO) alert. The ToO alert system will evaluate this alert and determine if the results warrant continued observations for the currently ongoing science case.

\section{The H.E.S.S. transients follow-up system in real-world applications}
\label{sec:realworld}
The real strength of the transients follow-up system originates from the interplay of the three subsystems described in the previous sections. The interplay can best be illustrated by looking at real-world use cases, which are commonly exploited by H.E.S.S. and other IACTs, and which highlight the core features of the system. While the entirety of the transients follow-up system functionalities are predefined, each science case uses a different subset of functionalities to allow for an optimal reaction of the telescope system. Follow-up observations of Gamma-Ray Bursts (GRBs) and Gravitational Wave (GW) events are two prominent examples for transients follow-up programs. Due to the short-lived nature of GRBs and GW events, a fast response of the full transients follow-up system is essential. Moreover, since the localisation accuracy of GW events is typically similar to or even larger than the H.E.S.S. FoV, an optimised pointing strategy of the H.E.S.S. telescopes needs to be employed. In the following, we will discuss the system response to a prompt GRB alert, and to a GW trigger. Applications to other science cases make use of either most or at least some of the components and steps described here.

\subsection{GRB follow-up observations}
The main use-case of the transients follow-up system is the prompt reaction to a short-lived transient, such as a GRB. In practise, this is how the system would respond to a GRB trigger issued by the X-ray satellite experiment \textit{Swift}-BAT~\citep{barthelmy2005burst}:

\begin{enumerate}
    \item The \textit{Swift}-BAT detects a GRB, relays the trigger to a ground station, from where a public alert is issued. This process typically takes a few tens of seconds.
    \item The ToO alert system receives the alert and initiates the alert processing.
    \item The ToO alert system tests all science cases that are applicable to a \textit{Swift}-BAT GRB alert. For simplicity, we here assume that two cases apply: if the GRB is immediately observable for H.E.S.S., a {\it prompt follow-up} is triggered, if the GRB is observable later and within a pre-defined time frame, an {\it afterglow follow-up} is considered. 
    \item The trigger criteria (e.g. source brightness and distance, maximum observation delay, zenith angle, sky brightness) are evaluated by the ToO alert system.
    \item The observation window is calculated. We assume that under the pre-defined conditions, the GRB position is visible immediately and for at least 5 minutes from the H.E.S.S. site.
    \item All relevant alert parameters are sent to the DAQ for scheduling. The operators are informed about the GRB alert by sounds and pop-up windows. Supplementary GRB alert information is distributed by E-mail to PIs and experts.
    \item The target is inserted into the current schedule for immediate observations. The current observations are stopped and removed from the schedule. Based on the ToO target parameters, the telescopes and RTA are configured.
    \item The fastest slewing path for CT5 is determined and used for CT5. The observations and data recording starts as soon as the target enters the CT5 FoV. CT1--4 join the observation once they slewed to the GRB position.
    \item Air shower events are recorded by the DAQ, camera images are calibrated by the RTA and reconstructed to derive main shower event properties.
    \item Gamma-ray like events are selected and filled into a sky image that is constantly updated and shown to the operators. The RTA checks if significant gamma-ray emission is detected from the GRB position and alerts the operators if certain significance thresholds are exceeded.
    \item The operators are monitoring the RTA results and contact the GRB expert on call to discuss further optimisations of the follow-up observations.
    \item The RTA results are written into a database for offline usage.
\end{enumerate}

Often, the location of a GRB is not visible right away ({\it afterglow follow-up}). In these cases, the sequence of events differs at point 5:
\begin{enumerate}
    \setcounter{enumi}{4}
    \item The observation window is calculated. The position is visible starting in e.g. 2 hours.
    \item The operators are alerted by sound and pop-up windows. Supplementary information is distributed by E-mail to PIs and experts.
    \item The GRB expert on call reviews the follow-up opportunity, checks regularly for additional MWL information, and decides whether or not to carry out the observations as recommended by the ToO alert system. If an alert retraction is received before the start of observations, the expert cancels the follow-up.
    \item As the start of the observation window approaches, operators schedule and initiate follow-up observations by running the script provided by the ToO alert system.
\end{enumerate}
From here on, the sequence continues as in the case of a reaction to prompt alerts from point 9. Over the course of the following day, additional MWL information and RTA results are assessed by the GRB expert on call and a decision for or against a continuation of the follow-up campaign is made. 

\subsection{Handling of alert updates}
The localisation uncertainty of \textit{Swift}-BAT-detected GRBs is typically much smaller than the FoV of H.E.S.S. Updated information for such alerts hence rarely changes the decision and implementation of H.E.S.S. observations on the timescale of the early GRB afterglow emission and the subsequent follow-up. However, alert information of instruments with poorer localisation capabilities are usually issued in a more complex way, which requires the handling of several corner cases.

\textit{Fermi}-GBM GRB alerts, for example, are typically issued in a sequence of alerts. The sequence can contain several {\it ground position} alert updates with changing reconstructed GRB positions. A final position estimate notice will not be generated for every GRB detected by the GBM. The localisation uncertainty of GBM-detected GRBs is often as large as, or even larger than, the H.E.S.S. FoV. The target coordinates may be updated a few times in quick succession, in some cases even outside the FoV that H.E.S.S. is currently observing. A potential counterpart can therefore emerge anywhere in the FoV. As described in the individual subsystem sections, the H.E.S.S. follow-up system is well equipped to address these cases of receiving sequences of alerts with updated coordinates. Naturally, it is possible, that the same event is detected by the \textit{Swift}-BAT, which typically provides a smaller localisation uncertainty. These complications are mitigated through a higher prioritisation of \textit{Swift} GRB alerts.

\subsection{Gravitational Wave follow-up observations}
The follow-up of GW triggers with H.E.S.S. is realised in a similar way as the GRB follow-up in that prompt and afterglow observations are typically initiated. However, another level of complexity needs to be handled by the system, namely the provisioning and handling of 2D and/or 3D uncertainty maps by e.g. LIGO/Virgo. As the localisation region is often of $\mathcal{O}(100\,\mathrm{deg}^{2})$ large, an optimised pattern of ordered telescope pointing positions needs to be defined. The modular and flexible design of the H.E.S.S. transients follow-up system allows for the easy integration of tailored and optimised algorithms to find the optimum pointing pattern. In the case of GW alerts, the transients follow-up system correlates the 3D localisation information with publicly available galaxy catalogues to maximise the chances to follow-up on potential counterparts. The detailed design, implementation and performance of this algorithm is described in \citet{Ashkar_2021}. This algorithm has proven its capabilities in the follow-up of GW170817, for which the first observation region covered the later-confirmed binary neutron star merger position in the galaxy NGC\,4993 \citep{abbott2017gw170817, abdalla2017gw170817}. During the GW observations, a counterpart in the RTA can emerge anywhere in the FoV of each of the scheduled observation regions. An automatic search of the entire RTA FoV and subsequent schedule adjustments based on the location of significant detection are not yet implemented (see also Sec.~\ref{sec:RTA}), so manual intervention by the operators is required to continue observations in this area.

GW alerts can also be sent out in sequences with updates to the localisation. Each alert in the sequence can change the optimised pointing pattern that should be followed. As GW alerts are followed up by a large number of instruments and observatories, the MWL information distributed in the community is manifold and can contain a much better localised counterpart candidate. As this information is, however, often not disseminated in machine-readable format, it makes an automatic handling of updated MWL information in the H.E.S.S. transients follow-up system challenging. At least for the moment, it requires experts to manage this information and instruct the H.E.S.S. telescope operators accordingly.

For all alerts that H.E.S.S. receives and that are later retracted by the issuing instrument, operators are informed and H.E.S.S. observations are stopped. A future extension of the GW follow-up algorithm could include updates of the pointing positions based on already observed FoVs and updates to the localisation.

\subsection{Validation and Maintenance}
With new science cases being regularly added to the portfolio, the variety of targets and their different follow-up strategies, the transients follow-up system has to be continuously validated and tested. This is particularly true for the interplay of the three subsystems. For this purpose, end-to-end system tests (so-called {\it fire drills}) are executed on a regular basis. In a fire drill, an alert with randomised coordinates is generated and sent to the ToO alert system. The tests are configured such that these self-generated alerts trigger observations for 5 minutes, allowing to validate all stages of the follow-up: receiving an alert, matching the science case, validating the observability, handover of the alert parameters to the DAQ, changing the observation schedule, repointing of the telescopes and starting of the observations, and receiving high-level results through the RTA.

Such tests are executed at least once per observing period under changing conditions to verify the systems integrity in (a combination of) various states:
\begin{enumerate}
\item At the end of a regular observing run, during run transitions or calibration runs,
\item with telescopes in a single or in multiple sub-arrays,
\item during astronomical darkness or during moderate moonlight,
\item with only a subset of the telescopes performing observations.
\end{enumerate}

In addition to the tests in the production system, an off-site test setup is available and running on a scaled-down version of the on-site infrastructure \citep{Zhu2021_DAQCluster}. This includes computing and storage nodes, switches, as well as the firewall. Key components like the RTA, VO system and DAQ ToO system with reduced functionality provide a development environment to ingest test alerts for testing new ToO system features, or re-running the RTA on existing data and testing new RTA features. It allows developers to validate changes and the interface integrity before new versions are deployed on site.

The overall maintenance effort for the system is small, with a varying level of maintenance being allocated to the three sub-systems. In particular the ToO Alert System needs regular updates due to the implementation of new science cases, changes to existing configurations or, going beyond maintenance, new processing features being developed to allow for more complex triggering schemes. In addition, the interface to the different instruments and alert stream channels may need to be adapted to changes in the metadata and data model of the incoming alerts. The interfaces between the ToO alert system, the DAQ and the H.E.S.S. scheduler have been identified as critical, as any change in one of the subsystems can cause undesirable side effects in the other subsystems and lead to a failure of the follow-up. The installation of new H.E.S.S. hardware, e.g. new Cherenkov cameras, requires the implementation and testing of the correct instrument reaction to the desired behaviour. The RTA also has to be adapted whenever the operation modes of H.E.S.S. change, and/or new components are installed. For instance, the introduction of observations under moderate moonlight with lower camera gain settings required an update of the instrument response functions.

\section{Conclusions and Outlook}
\label{sec:summary}

The H.E.S.S. transients follow-up system, a versatile and flexible system for the observations of transient objects with H.E.S.S. following multi-wavelength and multi-messenger triggers, is composed of the ToO alert system, and ToO-specific functions of the DAQ and RTA that were added as an extension to the existing H.E.S.S. system. The design of the transients follow-up system was driven by the variety of the transient events, their duration (from seconds to hours and days) and the desired reaction of H.E.S.S. to these alerts. The division into different subsystems with clear interfaces was guided by their core functionalities allowing for flexible extension of features. For instance, adding a new prompt science case does not usually require any changes to the DAQ, the RTA or any of the interfaces. Considering and implementing corner cases (see Sec.~\ref{sec:daq}) from the beginning was key to limiting development work during operation. The full system has been in stable operations since 2017 and is an important cornerstone for the H.E.S.S. transients science observations.

\subsection{Recent Science Results}
The interplay of the different components guarantees an optimal behaviour of the H.E.S.S. transients follow-up system: the fastest possible reaction through automation, correct handling of the corner cases, and the RTA can guide the decision on whether to continue observations. The system has allowed H.E.S.S. to further develop long-standing ToO programs such as the search for GRBs (see e.g.~\citep{aharonian2009hess}) and to actively participate in many large multi-instrument follow-up campaigns, as well as to probe new terrain with many exciting results, including:

\begin{itemize}
\item GW~170817 (Gravitational Wave): rapid follow-up of the first binary neutron-star merger, with the later identified event in the FoV of the first pointing position~\citep{gcn_gw170817_hess, abbott2017gw170817, abdalla2017gw170817},

\item IC~170922A (High-Energy Neutrino): rapid follow-up and monitoring after an neutrino alert, spatially and temporally coincident with the flare of TXS~0506+056~\citep{ic170922_followup_hess_atel, icecube2018multimessenger},

\item GRB~190829A (Gamma-ray Burst): rapid follow-up of the GRB with detection in the RTA in two consecutive nights~\citep{grb190829a_atel, abdalla2021revealing},

\item SGR~1935+2154 (Soft gamma-ray Repeater): follow-up triggered through correlation of a Swift-BAT alert with a catalogue of SGR candidates~\citep{abdalla2021searching},

\item RS~Ophiuchi (Galactic Nova): follow-up with an RTA detection that guided the monitoring campaign~\citep{rsoph_detection_atel, rsoph_spectrum_atel, rsoph_science},

\item PKS\,0346$-$27 (High-redshift AGN): follow-up of the flaring AGN PKS\,0346$-$27 at a redshift of $\sim$1 in a monitoring campaign using feedback from the RTA~\citep{pks0346_atel}.
\end{itemize}

\subsection{Future Developments in the Field}
Many of the transients science cases implemented in the H.E.S.S. observation program are automatically monitored and processed by the ToO Alert System. In some cases, however, H.E.S.S. still relies on PIs to monitor services such as Astronomer's Telegram or GCN Circulars that provide observational results in non-machine-readable format, and then manually request ToO follow-up observations to be included in the observation schedule. More homogeneous and automated handling, and thus a faster response to alerts, could be achieved if more alerts were submitted in machine-readable format. For example, services such as the Transients Name Server (TNS)\footnote{https://www.wis-tns.org} provide machine-readable access to a variety of reports and measurements of transient events.
Speeding up the announcement of public alerts may help other facilities decide earlier whether they also want to follow-up a particular event. Improved and regular reporting by facilities on the start of the follow-up observations could also enable for more frequent multi-wavelength / multi-instrument coverage. This, of course, depends on the different observatories/instruments and their policies, and relies on strong international standards, as promoted through international astronomical organisations such as the International Astronomical Union (IAU\footnote{https://www.iau.org/}) and International Virtual Observatory Alliance (IVOA\footnote{https://www.ivoa.net/}), for the communication of the results and the sharing of the observation schedules\footnote{see e.g. the services build using the IVOA VOEvent~\citep{petroff2017voevent, allan2017voevent} and ObsLocTAP standards~\citep{ivoa_spec_obsloctap}}.

Complex calculations or follow-up algorithms such as the convolution of uncertainties in gravitational wave localisation with galaxy catalogues or the correlation of (sub-threshold) alerts are often realised individually in every experiment. However, international networks in which representatives of different instruments and infrastructures can contribute could reduce some of the effort, increase synergies, and provide centralised and optimised algorithms. Some infrastructures are already exploring such options by developing or supporting community brokers that take the full stream of variable objects and enrich or classify transients for public alerts \citep[e.g.][]{AMON2013}. 
Another aspect that needs improvement concerns the sharing of follow-up decisions and observation schedules. Rapidly sharing each instrument's decision on whether to follow up on certain alerts would allow for near real-time coordination of the follow-up strategies for multiple instruments. Ideally, the additional publication and broader sharing of schedules would help to increase the multi-wavelength coverage for follow-up observations whilst optimising the available and limited observation time of the instruments. For instance, H.E.S.S. can use ATOM for simultaneous optical observations of interesting objects. The scientific prospects for brokering follow-up requests between instruments are clear, but so are the challenges for the technical implementation, coordination, and policies.

\subsection{Multi-Wavelength Outlook}
The landscape of instruments and infrastructures that monitor the sky for variable objects will change dramatically in the coming years. New survey instruments will become operational for essentially every wavelength range, such as the Square Kilometer Array (SKA)~\citep{carilli2004science} in the radio band, the Rubin Observatory~\citep{rubin_observatory} in the optical wavelength range, or SVOM~\citep{svom} probing the X-ray sky. With the improved sensitivity of all these instruments, transients follow-up programs are confronted with an avalanche of alerts on newly discovered transient objects. Some estimates range as high as one million transients detections per night. With such alert rates, the pre-processing, classification and filtering of alerts will be an essential task in the network between survey and follow-up instruments. Community or service brokers will be mandatory to provide sub-streams of alerts for events classified as belonging to specific astrophysical object classes. Such sub-streams will play a vital role in reducing the rate of new alerts to a manageable level, also for instruments probing the VHE gamma-ray sky.

The next-generation Cherenkov Telescope Array (CTA) Observatory will probe the energy range from tens of GeV to \mbox{$\sim$100 TeV}~\citep{cta2018science} and is currently in the construction phase. The large number of telescopes will allow observations in multiple sub-arrays simultaneously --- a key factor in either speeding up the scanning of large uncertainty regions or following multiple alerts at the same time. Transients-handling functionality is foreseen both at the level of an individual array location~\citep{oya2019array} and at the observatory level to enable coordinated responses from the two planned sites, as well as pre-processing in complex follow-up cases. CTA will be operated as an open observatory, which will result in many new use cases for the CTA transients follow-up system throughout the years of operation --- just as the H.E.S.S. transients follow-up system is still continuously being extended in terms of functionality and science-case implementation.

\section*{Acknowledgements}
We would like to thank all members of the H.E.S.S. collaboration for their technical support, helpful discussions, and the quick adoption of the H.E.S.S. transients follow-up system which triggered growing demands and further development of new features. The frequent use and enthusiasm for new possibilities allowed for the continuous expansion of the systems features and made it the success that it is. We would also like to thank the referee, Daniela Dorner, for her very constructive feedback, which significantly improved the manuscript.

\bibliographystyle{aa} 
\bibliography{ToOPaper}

\begin{appendix}
\section{ToO alert system implementation}
\label{app:ToO}
\begin{figure*}[t!]
\begin{center}
    \resizebox{0.7\hsize}{!}{\includegraphics{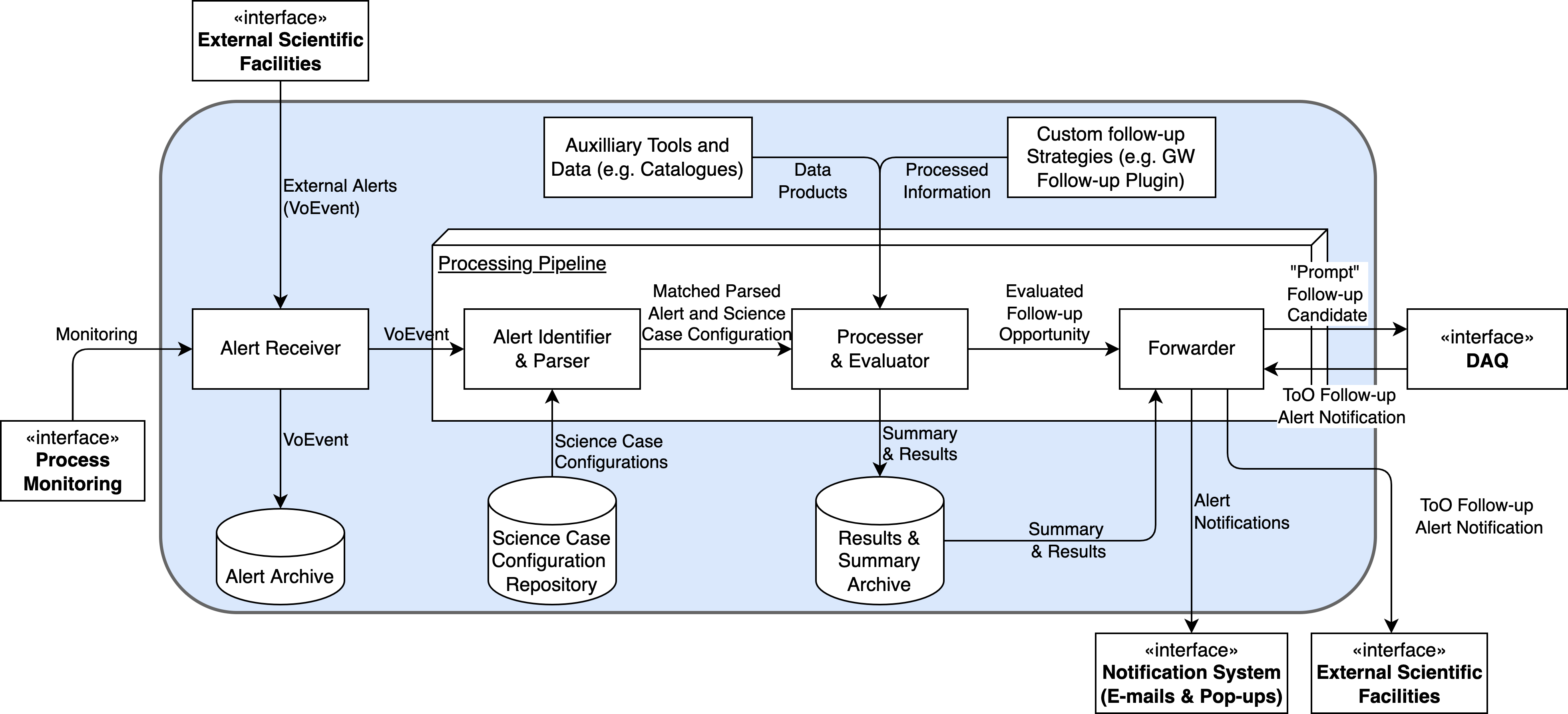}}
    \caption{Implementation view of the ToO Alert system. The {\it Receiving}, {\it Archiving}, and ToO Alert Scheduling Preparation functions map directly to components in the implementation. The {\it Processing Pipeline} implements the Matching of Alerts and Programs matching, the Alert Processing and the Scheduling Preparation. The Assessment of the Visibility function e.g. maps to the Processor \& Evaluator module in the pipeline.}
    \label{fig:vo_implementation_view}
\end{center}
\end{figure*}

The receiving and initial filtering of the alert is done in the \textit{Alert Receiver}, while a \textit{Processing Pipeline} implements the remaining tasks. Figure~\ref{fig:vo_implementation_view} shows the technical implementation of the ToO alert system.

\subsection{Alert Receiver}
The \textit{Alert Receiver} is implemented as an instance of the open-source {\it comet}~\citep{swinbank2014comet} software, which provides all needed functionality to broker VoEvent2.0~\citep{petroff2017voevent, allan2017voevent}, the current IVOA standard for time-domain alerts. We utilise the {\it voevent-parse} python package~\citep{staley2014voevent} to access the alert contents throughout the ToO alert system. A custom plugin to {\it comet} archives the alerts it receives and starts the processing pipeline as a new process. This way, the \textit{Alert Receiver} stays available for new alerts and multiple alerts can processed in parallel. {\it Comet} provides a number of useful functions that are used in the implementation of the \textit{Alert Receiver}, such as filtering duplicate alerts and checking against a list of trusted IPs that are allowed to send alerts. Received alerts are written into the alert archive and handed to the \textit{Processing Pipeline} for further analysis. The receiver archives all incoming alerts independent of the configured science cases, since the matching to the science cases happens only in the \textit{Processing Pipeline} where unsupported alerts are dropped.

The \textit{Alert Receiver} is configured with subscriptions to multiple data streams, among them the GCN, 4PiSky~\citep{staley20164} and the TAToO~\citep{ageron2012antares} system of the Mediterranean neutrino telescope Antares. Furthermore, a number of IPs are explicitly allowed for direct submission of alerts, such as the IPs from the IceCube computing clusters at the University of Wisconsin-Madison, which are running the IceCube real-time alert system~\citep{aartsen2017icecuberta}, and IPs from ATOM, the robotic optical telescope located at the H.E.S.S. site. 

\subsection{Processing Pipeline}
The \textit{Processing Pipeline} comprises a collection of science case configurations and modular processing scripts, and performs three main tasks: the matching of incoming alerts to science cases, the actual processing and evaluation of trigger conditions, and, finally, the preparation of the observation strategy and forwarding of the scheduling information of the follow-up candidate to the DAQ system, the operators and experts.

\FloatBarrier

\section{DAQ ToO system implementation}
\label{app:DAQ}
\begin{figure*}[t!]
\begin{center}
    \centering
    \resizebox{0.7\hsize}{!}{\includegraphics{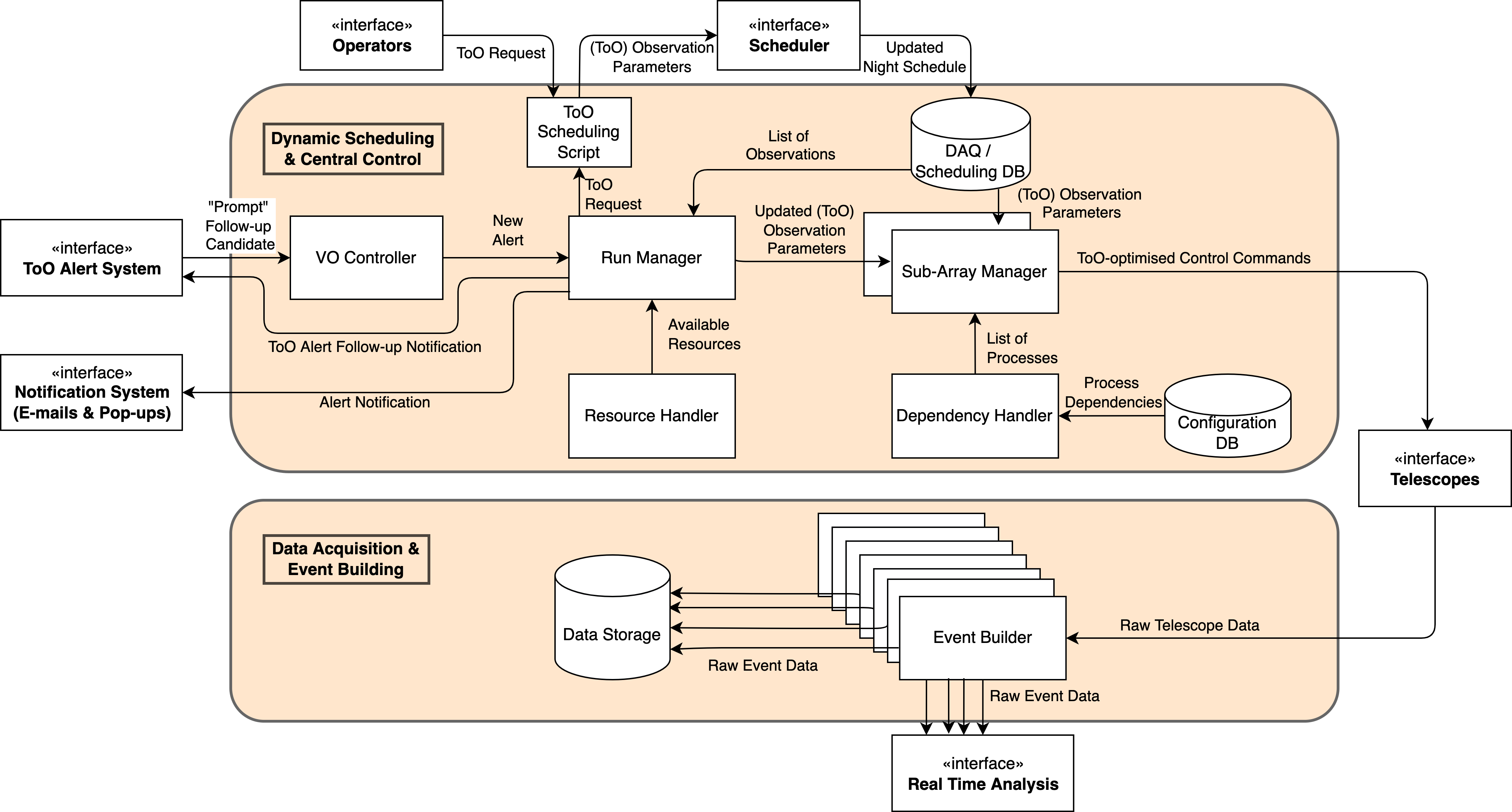}}
    \caption{Implementation view of the DAQ ToO system. The {\it Dynamic Scheduling} and {\it Central Control} functions are implemented in a common component. The {\it Run Manager} is the central component that takes input from the VO Controller, interacts with the Scheduler and initiates the Sub-Array Managers. The {\it Data Acquisition and Event Building} function is implemented in a  component that manages the many {\it Event Builders}.}
    \label{fig:daq_implementation_view}
\end{center}
\end{figure*}

Figure~\ref{fig:daq_implementation_view} shows the main components of the DAQ ToO system, the \textit{VO Controller} and the \textit{Run Manager} that were added or updated, respectively, to enable the fully automatic ToO follow-up.

\subsection{Dynamic Scheduling}
The \textit{VO Controller} (see Fig.~\ref{fig:daq_implementation_view}) connects the DAQ system to the ToO alert system. It receives and processes follow-up candidate alerts from the ToO alert system, checks their type and priority, queues them for later execution if needed and otherwise passes them to the \textit{Run Manager} for immediate execution. This mechanism allows queued alerts to be executed at a later stage in order of their priority. Currently, this only affects non-prompt alerts, since prompt alerts have the highest priority according to the H.E.S.S. policy and can interrupt ongoing observations. In the rare case where multiple updates of the follow-up candidate for the same underlying object are received, the \textit{VO Controller} grants a built-in grace period of a few tens of seconds for buffering and filtering before forwarding the final alert to the \textit{Run Manager}. This allows the already initiated stop-start cycle to be properly completed before the updates are carried out in one go rather than in several steps. In addition, the reaction to prompt follow-up alerts can be switched on and off via the \textit{VO Controller}, which is connected to a corresponding switch in the operator's user interface.
The \textit{Run Manager} steers the observations in the various telescope sub-arrays. It reads and holds the list of the scheduled and ongoing observations and is responsible for updating and distributing the observation parameters to each sub-array. For each incoming alert, the \textit{Run Manager} produces new or updated observation parameters depending on the type of the alert and distributes them together with the alert information to the \textit{Sub-Array Managers} that will further execute the observations. When it receives a prompt alert for a new ToO target from the \textit{VO Controller}, the \textit{Run Manager} calls the \textit{Scheduler} via a dedicated \textit{ToO Scheduling Script} to create a new schedule for the night using the alert parameters and the available telescopes tracked via the \textit{Resource Handler}. The updated schedule is stored in a scheduling database where it is available to the \textit{Run Manager} during execution. The \textit{Run Manager} notifies the operators using sound and pop-up windows with extended information on the alert and scheduling parameters in case of prompt or afterglow follow-up candidates. The \textit{ToO Scheduling Script} called by the \textit{Run Manager} is the same script that the operators can use to manually initiate delayed ToO observations following an afterglow alert based on the scheduling information given in the pop-up window. The \textit{Run Manager} also sends the relevant information to the RTA so that it can prepare a ToO-optimized analysis pipeline (see also Sec.~\ref{sec:RTA}) and informs the PIs and experts via E-mail, and, according to H.E.S.S. policies, external facilities (e.g. ATOM) via VOEvent messages using the \textit{Comet} infrastructure of the ToO alert system about the imminent follow-up observations.

\subsection{Central Control}
Before the observations start, the {\it Run Manager} checks the available telescopes and compares them with the telescopes specified in the distributed run parameters for the observations. If the request cannot be met, the observations are not started. Since (prompt) ToO observations have the highest priority and to achieve highest coverage, the current dedicated ToO observation mode requires that at least one telescope is available. After the start of the run, it is then up to the operators to decide whether ToO observations should continue, even if conditions are unfavourable (e.g. if only one of the CT1-4 telescopes is available). The dedicated ToO observation mode is part of the run parameters further distributed from the \textit{Run Manager} to the \textit{Sub-Array Managers} (see Fig.~\ref{fig:daq_implementation_view}). These are responsible for executing the observations in the respective group of telescopes and interact with the \textit{Controllers} of the telescopes, and their components (e.g. drive, Cherenkov camera) to perform the stop-start cycle and track their status during the observation execution. While the interactions between the telescope's drive and Cherenkov camera described above are handled directly by the respective \textit{Controllers}, it is the \textit{Sub-array Managers} that keep track of which components are part of the observations, what their status is and whether they are required or optional. The parameters of the ToO observation mode and the dependency of the components are easily configurable via databases, from which they are read by the \textit{Sub-Array Manager} or the supporting \textit{Dependency Handler} process. 

\section{RTA implementation}
\label{app:RTA}

\begin{figure*}[t!]
\begin{center}
\resizebox{0.7\hsize}{!}{\includegraphics{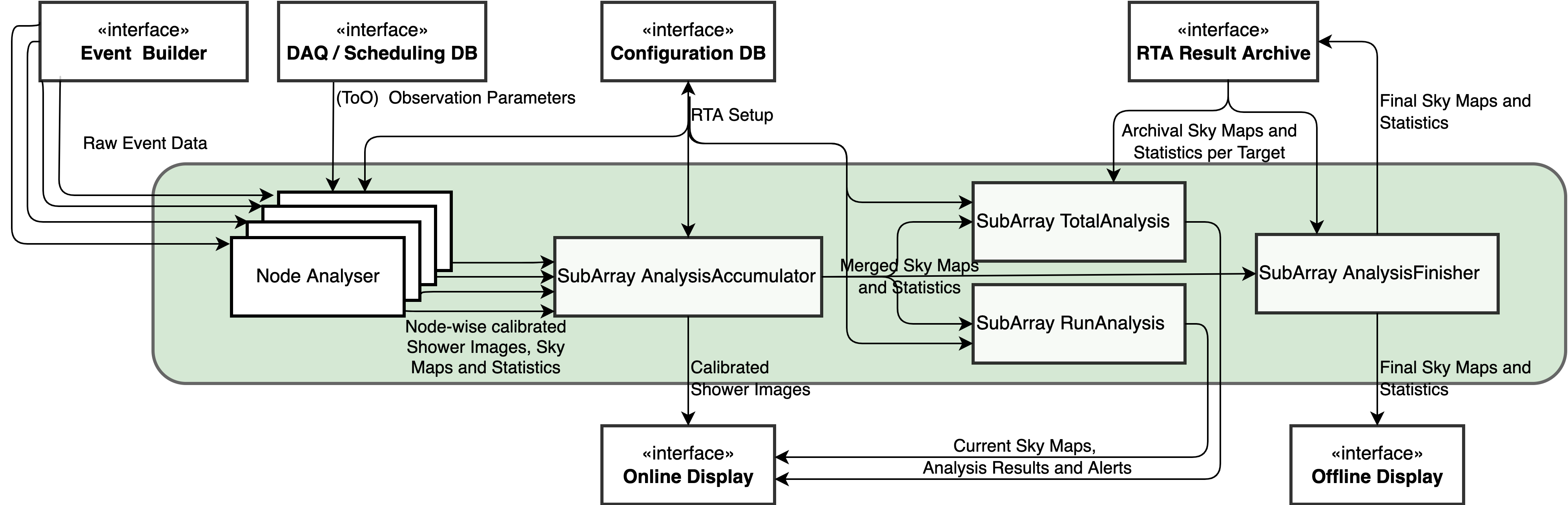}}
\caption{Implementation view of the RTA. The main functions of the RTA are implemented in a single Pipeline that manages many {\it Node Analysers} in order to receive input from the many {\it Event Builders} in the DAQ and Central Control system.}
\label{fig:rta_impl}
\end{center}
\end{figure*}

The RTA pipeline was designed to use as many processing components of the (long-established and versatile) H.E.S.S. software as possible and to only adapt methods where strictly necessary. This concerns all levels of the data analysis pipeline, from pixel calibration to event reconstruction and generation of high-level analysis products such as sky maps. The RTA implementation follows a modular approach to be able to a) distribute processing steps across computing nodes, and b) exchange each computing step with a dedicated real-time algorithm to improve speed.

Figure~\ref{fig:rta_impl} shows the implementation of the RTA in the on-site computing cluster, where it runs alongside the \textit{Event Builder} (see Sec.~\ref{sec:daq}). The RTA is operating on the distributed system of computing nodes that also perform the event building (see Sec.~\ref{sec:daq}). The stream of raw camera data, which is provided by one \textit{Event Builder} for 4 seconds before switching to the next \textit{Event Builder} on another computing node, enters the corresponding \textit{Analyser} process. This process is configurable by the \textit{Configuration DB} and the \textit{DAQ / Scheduling DB}, which support the operation of the RTA in different sub-arrays and analysis setups. For instance, for all observations that are marked as type \textit{ToO observations}, a CT5 mono analysis is set up, which allows for the lowest possible energy threshold, while sacrificing some performance of the array at higher energies. For regular observations, a hybrid analysis is initiated, which provides improved sensitivity in the core energy range of H.E.S.S. around 1\,TeV. Similar to the regular H.E.S.S. off-site analysis, a modular chain of software tasks is set up, in which each \textit{Analyser} process performs the event-wise pixel calibration, cleaning of shower images from NSB noise, the reconstruction of the air shower, and event classification. Gamma-ray candidate events that pass the signal/background separation are filled into sky maps.

The \textit{Analyser} process implements the \textit{Analysing events} stage in the RTA functionality diagram in Fig.~\ref{fig:rta_func}. Objects storing the cleaned shower images and node-wise sky maps and statistics at the target position are sent from the \textit{Analyser}, running on multiple computing nodes, to a single \textit{Accumulator} process that merges all the information from one sub-array.

This process is also configurable via the \textit{DAQ / Scheduling DB} and the \textit{Configuration DB}, which e.g. set the pace for sending camera images for events to the \textit{Online Display} that the operators monitor during data taking. The merged sky maps and statistics are sent to two separate processes: the \textit{RunAnalysis} and the \textit{TotalAnalysis} process. Both processes perform the background estimation in the field-of-view by applying the ring background technique to the merged sky maps as provided by the \textit{Accumulator} process. While the \textit{RunAnalysis} process merges intra-observation information, the \textit{TotalAnalysis} process also loads archival data from the \textit{RTA Results Archive} and the same field-of-view at the beginning of the observation and hence accumulates RTA results from more than one observation. The maximum look-back time is configurable and is by default set to twenty observations per target. Observations are typically conducted in 28-minute chunks on a single position. With the help of the \textit{Configuration DB}, the pace at which the \textit{RunAnalysis} and \textit{TotalAnalysis} processes calculate significance maps and derive plots of the squared angular distribution of event arrival directions with respect to the target position can be set. While the significance at the target position is accumulated continuously, the \textit{RunAnalysis} and \textit{TotalAnalysis} processes check if the significance at the target position exceeds configurable thresholds, in which case it alerts the operators with sounds and pop-up windows in the \textit{Online Display}. At the end of the observation, and when the transition to a new observation is started, an \textit{AnalysisFinisher} process is launched that produces final sky maps and statistics for the current observation, and accumulates archival data as described before. The \textit{RTA Result Archive} is filled with the observation result and the accumulated result for this field-of-view for further inspection. The \textit{AnalysisFinisher} process runs independently of the ongoing observation and therefore does not delay the initialisation and start of a new observation.

The on-site computing resources described in Sec.~\ref{sec:daq} allow for fast event building and calibration as well as shower reconstruction and classification. The RTA uses a dedicated real-time algorithm for the pixel intensity calculation, based on a running pedestal subtraction \citep{Funk2005}, and assumes fixed calibration coefficients (flatfielding, ADC-to-pe ratio, High-Low-ratio) for the pixel-to-pixel response. Although state-of-the-art pixel-wise likelihood analysis techniques such as \textit{ImPACT} or the \textit{model analysis} are not yet used on-site, other machine-learning techniques are used in the reconstruction and classification \citep{Ohm2009,Murach2015}.

\end{appendix}
\end{document}